\documentclass[prc,twocolumn,aps,superscriptaddress,showpacs]{revtex4}
\usepackage{amssymb}
\usepackage{amsmath,bm}
\usepackage{graphicx}
\usepackage[normalem]{ulem}
\usepackage[dvips]{color}
\usepackage{leftidx}
\usepackage{mhchem} 
\renewcommand{\sout}{\bgroup \color{red} \ULdepth=-.5ex \ULset}

\begin{document}

\title{Extended Skyrme interactions for nuclear matter, finite nuclei and  neutron stars}
\author{Zhen Zhang}
\affiliation{Department of Physics and Astronomy and Shanghai Key Laboratory for
Particle Physics and Cosmology, Shanghai Jiao Tong University, Shanghai 200240, China}
\author{Lie-Wen Chen\footnote{%
Corresponding author (email: lwchen$@$sjtu.edu.cn)}}
\affiliation{Department of Physics and Astronomy and Shanghai Key Laboratory for
Particle Physics and Cosmology, Shanghai Jiao Tong University, Shanghai 200240, China}
\affiliation{Center of Theoretical Nuclear Physics, National Laboratory of Heavy Ion
Accelerator, Lanzhou 730000, China}
\date{\today}

\begin{abstract}
Recent progress in theory, experiment and observation challenges the mean field models using
the conventional Skyrme interaction, suggesting that the extension of the conventional
Skyrme interaction is necessary.
In this work, by fitting the experimental data of a number of finite nuclei
together with a few additional constraints on nuclear matter using the simulated
annealing method,
we construct three Skyrme interaction parameter sets, namely, eMSL07, eMSL08 and eMSL09,
based on an extended Skyrme interaction which includes additional momentum and density dependent
two-body forces to effectively simulate the momentum dependence of the three-body force. The three
new interactions can reasonably describe the ground-state properties and the isoscalar
giant monopole resonance energies of
various spherical nuclei used in the fit as well as the ground-state properties of many other spherical nuclei,
nicely conform to the current knowledge on
the equation of state of asymmetric nuclear matter, eliminate the notorious unphysical instabilities
of symmetric nuclear matter and pure neutron matter up to a very high density of $1.2$ fm$^{-3}$, and
simultaneously support heavier neutron stars with mass larger than two times solar mass.
One important difference of the three new interactions is about the prediction of the symmetry energy
at supra-saturation densities, and these new interactions are thus potentially useful for the determination
of the largely uncertain high-density symmetry energy in future.
In addition, a comparison is made for the predictions of nuclear matter,
finite nuclei and neutron stars with the three new interactions versus those with three typical
interactions BSk22, BSk24 and BSk26 from Brussels group.
\end{abstract}

\maketitle

\section{Introduction}

The exact knowledge on infinite nuclear matter, which is intimately related
to the in-medium effective nuclear interactions, is of fundamental importance in
nuclear physics and astrophysics~\cite{Lat04,Ste05,Bar05,LCK08}. Especially,
the equation of state (EOS) of isospin asymmetric nuclear matter is the main
ingredient in the study of neutron stars.
In principle, the nuclear matter EOS can be obtained from various microscopic many-body
approaches, e.g., the nonrelativistic and relativistic Brueckner-Hartree-Fock (BHF)
method~\cite{Haa87,Bom91,Zuo99}, the variational many-body approach~\cite{Pan79,Akm98},
quantum Monte Carlo (QMC) method~\cite{Gez13,Wla14,Rog14} and chiral effective field theory
(ChEFT)~\cite{Heb10,Tew13}, using realistic nuclear forces. However, due to the poorly
known many-body interactions and the limitations in the techniques for solving the
nuclear many-body problem, accurate determination of properties of nuclear matter around
and beyond saturation density $\rho_0$ is still an open challenge for microscopic
many-body theory.
Another different perspective is the mean field model using phenomenological nuclear effective
interactions with several parameters adjusted by fitting experimental data~\cite{Ben03}.
Among various phenomenological interactions, the nonrelativistic zero-range density and
momentum-dependent Skyrme-type effective nucleon-nucleon interactions perhaps are
the most widely used.
Proposed by Skyrme in the 1950s~\cite{Sky56} and firstly applied in the study of finite
nuclei in Hartree-Fock (HF) calculations by Brink and Vautherin in 1970s~\cite{Vau72},
the Skyrme interaction enormously simplifies the calculations with its zero-range
form and has been very successfully used to describe the masses, charge radii and
excited states of finite nuclei as well as the EOS of nuclear matter around $\rho_0$.
Moreover, extrapolation to high-density region based on the Skyrme-Hartree-Fock (SHF) model
provides an important approach to investigate the properties of dense nuclear matter and
the relevant astrophysics problems, especially the properties of neutron stars.

Since 1970s, a lot of work has been devoted to improving the Skyrme interaction to
better reproduce the experimental data or describe different physical objects.
Recently, much attention has been given to the problem of searching
for effective interactions or energy density functionals (EDFs) which could
simultaneously reproduce the properties of nuclear matter and finite nuclei, and
at the same time be applicable for the study of neutron
stars~\cite{Cha97,Cha98,Ben03,Cao06,Sto07,Cha09,Gor10,Gam11,Dut12,Gor13,Bal13,Erl13,Che14,Gor15}.
This problem is of particular interest as two very heavy neutron stars with mass of
two times solar mass ($2M_{\odot}$) have been observed recently~\cite{Dem10,Ant13},
which requires the pressure of high-density nuclear matter should be large enough
to support such massive neutron stars against the strong gravity. This
requirement is a big challenge for many theoretical models and indeed
rules out essentially all the soft nuclear matter EOSs. Given the EOS of
symmetric nuclear matter has been relatively well constrained even up to
about $5\rho_0$ by analyzing the experimental
data on giant resonances of finite
nuclei~\cite{You99,Li07,Pat13} as well as the collective flows and
kaon production in heavy-ion collisions~\cite{Aic85,Dan02, Fuc06},
a softer symmetry energy at high densities
has been essentially ruled out by the observation of $2M_{\odot}$ neutron stars.
It should be pointed out that some new physical mechanisms,
such as non-Newtonian gravity~\cite{WenDH09,Kri09,Zha11,ZhengH12},
may support $2M_{\odot}$ neutron stars with a soft symmetry energy.
In the present work we shall focus on the standard nuclear physics
without considering these new physics.

During the last decade, considerable progress in determining the symmetry
energy or neutron matter EOS at subsaturation densities has been made,
both theoretically and experimentally. Indeed,
it has been well established that the binding energy of finite nuclei
can put rather stringent
constraints on the symmetry energy at a subsaturation density
$\rho \approx (2/3)\rho_0$~\cite{Hor01a,Fur02,Wan13,Dan14,Zha13,Bro13}.
For the symmetry energy at an even lower density $\rho \approx \rho_0/3$,
it has been shown recently~\cite{Zha15} that the measurement of the
electric dipole polarizability in $^{208}$Pb can give a quite accurate
constraint and the result is in very good agreement with the
constraints from SHF analyses of isobaric analog states and neutron
skin data~\cite{Dan14} as well as the transport model analyses of
mid-peripheral heavy-ion collisions of Sn isotopes~\cite{Tsa09}. For pure neutron matter,
very similarly, the binding energy per neutron around subsaturation
densities $(2/3)\rho_0$ and $\rho_0/3$ has been constrained recently by analyzing
the ground-state properties of doubly magic nuclei~\cite{Bro13} and the
electric dipole polarizability in $^{208}$Pb~\cite{Zha15}, respectively. In addition,
the calculations based on the microscopic ChEFT~\cite{Tew13} as well as the QMC
calculations~\cite{Gez13,Wla14,Rog14} have also provided very useful information
on the EOS of pure neutron matter, especially at subsaturation densities.
These theoretical and experimental constraints consistently favor a relatively
soft symmetry energy or EOS of asymmetric nuclear matter, at least at
subsaturation densities around $(2/3)\rho_0$. The symmetry energy softer at
subsaturation densities (favored by experimental constraints and theoretical
predictions) but stiffer at higher densities (favored by the observation of
$2M_{\odot}$ neutron stars ) challenges the SHF model with the conventional
Skyrme interactions. For example, the Skyrme interaction
TOV-min~\cite{Erl13}, which is built by fitting properties of both
finite nuclei and neutron stars, can successfully support $2M_{\odot}$ neutron
stars but predicts a neutron matter EOS significantly deviating from the
ChEFT calculations~\cite{Tew13} as well as the constraint extracted from analyzing the
electric dipole polarizability in $^{208}$Pb~\cite{Zha15} at densities below
about $0.5\rho_0$.

Furthermore, it is well known that a notorious shortcoming of the conventional
standard Skyrme interactions is that they predicts various instabilities of
nuclear matter around saturation density or at supra-saturation densities,
which in principle hinders the application of the Skyrme interactions in the
study of dense nuclear matter as well as neutron stars.
For instance, most of the conventional standard Skyrme interactions predict spin or spin-isospin
polarization in the density region of about
$(1\sim 3.5)\rho_0$~\cite{Mar02,Dut12}, including the
famous SLy4 interaction~\cite{Cha98} which has been widely used in both nuclear
physics and neutron star studies and leads to spin-isospin instability of symmetric nuclear matter
at densities beyond about 2$\rho_0$~\cite{Cha10}.
On the other hand,
the calculations based on the microscopic many-body theory using realistic nuclear
forces, such as relativistic and nonrelativistic BHF approach~\cite{Vid02,Sam07},
the QMC method~\cite{Fan01}, the ChEFT method~\cite{Sam15}, and the lowest order
constrained variational approach~\cite{Bor08}, predict no such instabilities at
densities up to substantially high densities.
To solve this problem, Margueron
and Sagawa proposed an extended form of the Skyrme interaction with
additional density-dependent terms~\cite{Mar09}, while Chamel \textit{et al.}
introduced momentum and density dependent terms which are density-dependent
generalizations of the usual $t_1$ and $t_2$ terms in the conventional
standard Skyrme interaction~\cite{Cha09}.
In Ref.~\cite{Cha10}, Chamel and Goriely find that the spin and spin-isospin
instabilities can be removed by omitting the time-odd terms in
$(\bm{s}_n+\bm{s}_p)\times(\bm{T}_n+\bm{T}_p)$ and
$(\bm{s}_n-\bm{s}_p)\times(\bm{T}_n-\bm{T}_p)$, namely setting
$C_0^T=C_1^T=0$ in the notation of Ref.~\cite{Ben02}.

In Ref.~\cite{Dut12}, Dutra \textit{et al.} find that, of 240
standard Skyrme interactions, only 6 satisfy all eleven constraints
on the properties of nuclear matter selected therein. However,
we would like to point out that all the $6$ Skyrme interactions
predict spin or spin-isospin instabilities below $3\rho_0$ and fail
to produce $2M_{\odot}$ neutron stars.
In addition, the Skyrme interactions constrained only from
nuclear matter properties usually have no common ability to reproduce
properties of finite nuclei~\cite{Ste13}.
A feasible approach to address the above problems existed in
the conventional standard Skyrme interactions is to extend the Skyrme interaction
by including additional terms to effectively simulate the
momentum dependence of the three-body force~\cite{Kre77,Ge86,Zhu88,Cha09,Gor10,Gor13,Gor15}.
In this work, based on the extended Skyrme interaction, we use the
simulated annealing method~\cite{Agr05} to
construct three Skyrme interaction parameter sets, namely, eMSL07, eMSL08 and eMSL09,
by fitting the experimental
data of ground-state properties and isoscalar giant monopole resonance (ISGMR) energies of some
finite nuclei together with a few additional constraints on properties of nuclear matter.
Our main purpose here is
to construct parameter sets of the extended Skyrme interaction
that can
reasonably describe the properties of finite (closed-shell or semi-closed-shell) nuclei, satisfy the
most recent constraints on nuclear matter, especially the EOS of asymmetric
nuclear matter at subsaturation densities,  eliminate the unphysical
instabilities of nuclear matter in the density region encountered in
neutron stars, and successfully support $2M_{\odot}$ neutron stars.
In addition,
a comparison is made for the predictions of nuclear matter,
finite nuclei and neutron stars with the new interactions constructed in the
present work versus those with the accurately calibrated interactions
BSk22, BSk24 and BSk26~\cite{Gor13} from
Brussels group.
Although constructed with a different strategy from that of the Brussels group,
the new extended Skyrme interactions
obtained in the present work give predictions for nuclear matter and neutron stars
that are very similar with the calculations using the extended Skyrme interactions
in BSk family~\cite{Gor13}, which are constructed mainly for accurately describing
the nuclear mass.
Furthermore, although the three new interactions constructed in
the present work give very similar predictions of the properties of finite nuclei
as well as nuclear matter at subsaturation densities, they predict
different supra-saturation density behaviors of the symmetry energy,
and these new interactions are thus potentially useful
for the determination of the largely uncertain high-density symmetry energy in future.

The paper is organized as follows. In Sec.~\ref{Sec:ESHF}, we introduce
the form of the extended Skyrme interaction and the energy density functional
adopted in this work. In Sec.~\ref{Sec:Fit}, the experimental data and
constraints used in our fitting are presented. The parameter sets of three
new extended Skyrme interactions and the corresponding results are exhibited
in Sec.~\ref{Sec:Result}. Finally, we summarize our conclusions in Sec~\ref{Sec:Summary}.
Appendix~\ref{App:EOSESHF} gives the expressions for several macroscopic
quantities in the extended SHF model and Appendix~\ref{App:MSL} presents
the analytical relations between the chosen macroscopic quantities and the
microscopic Skyrme parameters.

\section{Models and methods}
\label{Sec:ESHF}

\subsection{Extended Skyrme-Hartree-Fock model}

In the conventional SHF model, nucleons generally interact with each
other through the so-called standard Skyrme interaction
(see, e.g., Ref.~\cite{Cha97})
\begin{eqnarray}
v(\bm{r}_1,\bm{r}_2)&=&t_0(1+x_0P_{\sigma})\delta(\bm{r}_1-\bm{r}_2)   \notag \\
& &+\frac{1}{2}t_1(1+x_1P_{\sigma})[\bm{k}'^2\delta(\bm{r}_1-\bm{r}_2)+\mathrm{c.c.}] \notag \\
& &+t_2(1+x_2P_{\sigma})\bm{k}'\cdot\delta(\bm{r}_1-\bm{r}_2)\bm{k} \notag \\
& &+\frac{1}{6}t_3(1+x_3P_{\sigma})\rho ^{\alpha}\left(\frac{\bm{r}_1+\bm{r}_2}{2}\right)\delta(\bm{r}_1-\bm{r}_2)\notag\\
& &+iW_0(\bm{\sigma}_1+\bm{\sigma}_2)\cdot[\bm{k}'\times\delta(\bm{r}_1-\bm{r}_2)\bm{k}],
\label{Eq:Sky}
\end{eqnarray}
where $\bm{\sigma}_i$ is the Pauli spin operator, $P_{\sigma}$
is the spin-exchange operator,
$\bm{k}=-i(\bm{ \nabla}_1-\bm{\nabla}_2)/2$ is the relative
momentum operator, and
$\bm{k}^{\prime}$ is the conjugate operator of $\bm{k}$ acting
on the left.

In the present work, to effectively take account of the momentum
dependence of the three body interaction, we use the extended Skyrme
interaction with the following additional zero-range density-
and momentum-dependent terms~\cite{Kre77,Ge86,Zhu88,Cha09,Gor10,Gor13,Gor15}
\begin{eqnarray}
&+&\frac{1}{2}t_4(1+x_4P_{\sigma})
\left[\bm{k}'^2\rho^{\beta}\left(\dfrac{\bm{r}_1+\bm{r}_2}{2}\right)\delta(\bm{r}_1-\bm{r}_2)+\mathrm{c.c.} \right]\notag\\
&+&t_5(1+x_5P_{\sigma})\bm{k}'\cdot\rho^{\gamma}\left(\dfrac{\bm{r}_1+\bm{r}_2}{2}\right)\delta(\bm{r}_1-\bm{r}_2)\bm{k},
\label{Eq:ExSky}
\end{eqnarray}
which are just the density dependent generalization of the $t_1$ and
$t_2$ terms in Eq.~(\ref{Eq:Sky}). For simplicity, $\beta$ and $\gamma$ are
set to be unity in the present work, just like the form used in
Ref.~\cite{Kre77,Ge86,Zhu88}. It should be noted that in HF calculations
the zero-range momentum dependent three-body force is equivalent to a
momentum and density dependent two-body force for spin-saturated systems,
and the original values of $\beta$ and $\gamma$ are just unity~\cite{War79}.
Therefore, there are thirteen adjustable Skyrme parameters $t_0 \sim t_5$,
$x_0 \sim x_5$ and $\alpha$ in the present extended Skyrme interaction.

In Hartree-Fock approach with the extended Skyrme interaction, the total energy density  of a nucleus can be expressed as
\begin{equation}
\label{Eq:HESkyrme}
\mathcal{H}=\mathcal{K}+\mathcal{H}_0+\mathcal{H}_3+\mathcal{H}_{\mathrm{eff}}
+\mathcal{H}_{\mathrm{fin}}+\mathcal{H}_{\mathrm{SO}}
+\mathcal{H}_{\mathrm{sg}}+\mathcal{H}_{\mathrm{Coul}},
\end{equation}
where $\mathcal{K}=\frac{\hbar^2}{2m}\tau$ is the kinetic-energy
term and $\mathcal{H}_0$, $\mathcal{H}_3$, $\mathcal{H}_{\mathrm{eff}}$
, $\mathcal{H}_{\mathrm{fin}}$, $\mathcal{H}_{\mathrm{SO}}$, $\mathcal{H}_{\mathrm{sg}}$ are given by~\cite{Cha09}
\begin{eqnarray}
\mathcal{H}_0&=&\frac{1}{4}t_0[(2+x_0)\rho^2-(2x_0+1)(\rho_n^2+\rho_p^2)],  \label{Eq:H0} \\
\mathcal{H}_3&=&\frac{1}{24}t_3\rho^{\alpha}[(2+x_3)\rho^2-(2x_3+1)(\rho_n^2+\rho_p^2)], \label{Eq:H3} \\
\mathcal{H}_{\mathrm{eff}}&=&\frac{1}{8}[t_1(2+x_1)+t_2(2+x_2)]\rho \tau \notag \\
&-&\frac{1}{8}[t_1(2x_1+1)-t_2(2x_2+1)](\rho_n\tau_n+\rho_p\tau_p), \notag\\
&+&\frac{1}{8}[t_4(2+x_4)\rho^{\beta}+t_5(2+x_5) \rho^{\gamma}] \rho\tau \label{Eq:HEFF} \\
&-&\frac{1}{8}[t_4(2x_4+1)\rho^{\beta}-t_5(2x_5+1)\rho^{\gamma}](\rho_n\tau_n+\rho_p\tau_p),
\notag
\end{eqnarray}
\begin{eqnarray}
\mathcal{H}_{\mathrm{fin}}&=&\frac{1}{32}[3t_1(2+x_1)-t_2(2+x_2)]
(\nabla \rho)^2\notag \\
&-&\frac{1}{32}[3t_1(2x_1+1)+t_2(2x_2+1)]
\sum_{q=n,p}(\nabla \rho_q)^2\notag \\
&+&\frac{1}{32}[(2\beta+3)t_4(2+x_4)\rho ^{\beta}-t_5(2+x_5)\rho ^{\gamma}]
(\nabla \rho)^2 \notag\\
&-&\frac{1}{32}[3t_4(2x_4+1)\rho ^{\beta}+t_5(2x_5+1)\rho ^{\gamma}]
\sum_{q=n,p}(\nabla \rho_q)^2\notag \\
&-&\frac{\beta}{16}t_4(2x_4+1)\rho ^{\beta-1}\nabla \rho \sum_{q=n,p}\rho_q\nabla\rho_q,
\end{eqnarray}
\begin{eqnarray}
\mathcal{H}_{\mathrm{SO}}&=&\frac{1}{2}W_0[J\cdot\nabla \rho+J_n\nabla \rho_n+J_p\nabla\rho],
\end{eqnarray}
\begin{eqnarray}
\mathcal{H}_{\mathrm{sg}}&=&-\frac{1}{16}(t_1x_1+t_2x_2)J^2+\frac{1}{16}(t_1-t_2)\sum_{q=n,p}J_q^2 \notag \\
&-& \frac{1}{16}(t_4x_4\rho^{\beta}+t_5x_5\rho^{\gamma})J^2 \notag \\
&+& \frac{1}{16}(t_4\rho^{\beta}-t_5\rho^{\gamma})\sum_{q=n,p}J_q^2.
\end{eqnarray}
Here, $\rho=\rho_n+\rho_p$, $\tau =\tau_n+\tau_p$, and $J = J_p + J_n$ are the
particle number density, kinetic-energy density, and spin
density, with $p$ and $n$ denoting the protons and neutrons,
respectively. The Coulomb energy density can be expressed as
\begin{equation}
\mathcal{H}_{\mathrm{Coul}}=\mathcal{H}_{\mathrm{Coul}}^{\mathrm{dir}}
+\mathcal{H}_{\mathrm{Coul}}^{\mathrm{exc}},
\end{equation}
where $\mathcal{H}_{\mathrm{Coul}}^{\mathrm{dir}}$ is the direct term
of the form
\begin{equation}
\mathcal{H}_{\mathrm{Coul}}^{\mathrm{dir}}=\frac{1}{2}e^2\rho_p(r)
\int\frac{\rho_p(r^{\prime})d^3r^{\prime}}{\vert \bm{r}-\bm{r}^{\prime}\vert},
\end{equation}
and $\mathcal{H}_{\mathrm{Coul}}^{\mathrm{exc}}$ is the exchange term
\begin{equation}
\mathcal{H}_{\mathrm{Coul}}^{\mathrm{exc}}(r)=-\frac{3}{4}e^2\rho_p(r)
\left( \frac{3\rho_p(r)}{\pi} \right)^{1/3}.
\end{equation}

In the present work, we include the center-of-mass correction
in the first order to the binding energy by modifying nucleon
mass $m$ to $mA/(A-1)$ with $A$ the nucleon number.
The pairing energy is evaluated in the constant gap approximation
with the gap ~\cite{Rin80}
\begin{equation}
\Delta =\frac{11.2}{\sqrt{A}}\mathrm{MeV}.
\end{equation}
We would also like to point out that the contributions of spin current
tensor terms $J^2$ and $J_q^2$ are also included in our calculations.

\subsection{Skyrme parameters and macroscopic quantities of nuclear matter}

The expression  Eq.~(\ref{Eq:HESkyrme}) can be rewritten as
\begin{eqnarray}
\mathcal{H}&=&  \mathcal{K}+\mathcal{H}_0+\mathcal{H}_3+\mathcal{H}_{\mathrm{eff}}
+\frac{G_S}{2}(\nabla \rho)^2-\frac{G_V}{2}(\nabla \rho_1)^2  \notag \\
&~&-\frac{G_{SV}}{2}\delta\nabla\rho\nabla \rho_1
+\mathcal{H}_{\mathrm{Coul}}+\mathcal{H}_{\mathrm{SO}}+\mathcal{H}_{\mathrm{sg}},
\end{eqnarray}
where
$G_S$ is the gradient coefficient, $G_V$ is the symmetry-gradient
coefficient, $G_{SV}$ is the cross gradient coefficient, and $\delta=\rho_1/\rho$
is the isospin asymmetry with $\rho_1 =\rho_n-\rho_p$.
In the limit of infinite static nuclear matter, the sum of
$\mathcal{K}+\mathcal{H}_0+\mathcal{H}_3+\mathcal{H}_{\mathrm{eff}}$
corresponds to the nuclear matter energy density $\rho E(\rho, \delta)$ where
$E(\rho, \delta)$ is just the EOS of asymmetric nuclear matter.
Conventionally, some macroscopic quantities are introduced to
characterize the EOS of asymmetric nuclear matter. For example,
the $E(\rho, \delta)$ can be expanded as
\begin{eqnarray}
E(\rho,\delta) = E_0(\rho) + E_{\mathrm{sym}}(\rho)\delta^2+\mathcal{O}(\delta^4),
\end{eqnarray}
where $E_0(\rho)$ is the EOS of symmetric nuclear matter and
$E_{\mathrm{sym}}(\rho)$ is the symmetry energy.
The $E_0(\rho)$ is usually expanded around saturation density
$\rho_0$ in terms of the incompressibility coefficient $K_0$ and
the skewness coefficient $J_0$ as
\begin{eqnarray}
E_0 = E_0(\rho_0) + \frac{1}{2!}K_0\chi^2+\frac{1}{3!}J_0\chi^3+\mathcal{O}(\chi^4),
\end{eqnarray}
with $\chi =(\rho-\rho_0)/{(3\rho_0)}$. Similarly, the
$E_{\mathrm{sym}}(\rho)$ can be expanded around a reference density
$\rho_r$ in terms of the density slope parameter $L$ and
the density curvature parameter $K_{\mathrm{sym}}$ as
\begin{eqnarray}
E_{\mathrm{sym}}(\rho) &=& E_{\mathrm{sym}}(\rho_r) + L(\rho_r) \chi_r   \notag \\
&+& \frac{1}{2!}K_{\mathrm{sym}}(\rho_r)\chi_r^2+\mathcal{O}(\chi_r^3),
\end{eqnarray}
with $\chi_r=(\rho-\rho_r)/(3\rho_r)$.
The $E_0(\rho_0)$, $K_0$, $J_0$,
$E_{\mathrm{sym}}(\rho_r)$, $L(\rho_r)$ and $K_{\mathrm{sym}}(\rho_r)$
are six important macroscopic quantities that characterize
the EOS of asymmetric nuclear matter.

According to the analysis method in the modified Skyrme-like (MSL) model
with the standard Skyrme interaction~\cite{Che09,Che10}, the nine Skyrme
parameters $t_0\sim t_3$, $x_0 \sim x_3$ and $\alpha$ are expressed
explicitly in terms of nine macroscopic quantities $\rho_0$, $E_0(\rho_0)$, $K_0$,
$E_{\mathrm{sym}}(\rho_r)$, $L(\rho_r)$, $G_S$, $G_V$, the isoscalar
effective mass at saturation density $m_{s,0}^{\ast}$ and isovector effective mass at saturation density
$m_{v,0}^{\ast}$. For the extended Skyrme interaction
with thirteen Skyrme parameters in the present work, four additional
macroscopic quantities are needed to determine the thirteen Skyrme parameters, and here we use the skewness coefficient
$J_0$, the density curvature parameter of the symmetry energy
$K_{\mathrm{sym}}(\rho_r)$, the cross gradient coefficient
$G_{SV}$ and the Landau parameter $G_0^{\prime}(\rho_0)$ of symmetric nuclear matter in the
spin-isospin channel. Here $J_0$ and $K_{\mathrm{sym}}(\rho_r)$, two
higher-order quantities in nuclear matter EOS, may have important
impacts on neutron star properties but are still poorly known~\cite{Che09,ChenLW11,Cai14}.
The coefficient $G_{SV}$ vanishes in the standard SHF model. The Landau
parameter $G_0^{\prime}$ determines, to the leading order, the spin-isospin
properties of nuclear matter and its value at saturation density can
vary from about $0$ to $1.6$ depending on the models and
methods~\cite{Agr05,Ost92,Bal98,Ben02,Zuo03,She03,Wak05,Bor06}.
In Appendix~\ref{App:EOSESHF}, we present the explicit expressions of several macroscopic quantities
in the SHF model with the extended Skyrme interactions.
And the analytical relations between the thirteen macroscopic
quantities, i.e., $\rho_0$, $E_0(\rho_0)$, $K_0$, $J_0$,
$E_{\mathrm{sym}}(\rho_r)$, $L(\rho_r)$, $K_{\mathrm{sym}}(\rho_r)$,
$G_S$, $G_V$, $G_{SV}$, $m_{s,0}^{\ast}$, $m_{v,0}^{\ast}$ and
$G_0^{\prime}$, and the thirteen Skyrme parameters
$t_0 \sim t_5$, $x_0 \sim x_5$ and $\alpha$ with fixed $\beta$ and
$\gamma$ can be found in Appendix~\ref{App:MSL}.

The MSL method, in which all the Skyrme parameters are expressed in
terms of macroscopic quantities, provides a simple and convenient approach to consider
theoretical, experimental or empirical constraints on the selected
macroscopic quantities for the properties of asymmetric nuclear matter
in the SHF calculations.
Another important advantage of the MSL method is
that one can easily examine the correlation of experimental
data or observations with each individual macroscopic quantities
by varying individually these macroscopic quantities within
their empirical ranges~\cite{Che10}. In the present work, instead
of making correlation analysis, we shall focus on building parameter
sets of the extended Skyrme interactions.

\subsection{Landau parameters}
\label{Chap:LP}
The stability of nuclear matter can be investigated using
the Landau Fermi-liquid theory. In this approach, for symmetric
nuclear matter, the interaction $V(\bm{k},\bm{k}^{\prime})$
between two quasiparticles at Fermi surface with momentum $\bm{k}$ and $\bm{k}^{\prime}$
is obtained from
a second-order variation of the energy density $\mathcal{E}$ with
respect to a variation of distribution function of the quasiparticles,
and it can be usually written as
\begin{eqnarray}
\label{Eq:LP}
V(\bm{k},\bm{k}^{\prime})&=&\delta(\bm{r})N_0^{-1} \sum_{l}
\left[ F_l+F_l^{\prime} \bm{\tau}_i\cdot \bm{\tau}_j +G_l \bm{\sigma}_i\cdot \bm{\sigma}_j  \right.  \notag \\
&+&\left. G_l^{\prime} \left( \bm{\tau}_i\cdot \bm{\tau}_j  \right)
 \left( \bm{\sigma}_i\cdot \bm{\sigma}_j \right)\right]P_l(\mathrm{cos}\theta),
\end{eqnarray}
where $P_l$ is the Legendre polynomial, $\theta$ is the angle between
$\bm{k}$ and $\bm{k}^{\prime}$, and $N_0$ is the level density at the
Fermi surface defined as
\begin{equation}
N_0=\frac{2m_s^{\ast}k_F}{\hbar^2\pi^2},
\end{equation}
where $k_F=(3\pi^2\rho/2)^{1/3}$ is the Fermi momentum and $m_s^{\ast}$
is the isoscalar effective mass at density $\rho$. $F_l$, $F_l^{\prime}$, $G_l$
and $G_l^{\prime}$ are the so-called dimensionless Landau parameters.
For the Skyrme interaction containing only $S$ and $P$ wave contributions
that we are considering in the present work, all Landau parameters
vanish for $l>1$. Therefore, there are eight Landau parameters for symmetric nuclear matter,
i.e., $F_l$, $F_l^{\prime}$, $G_l$ and $G_l^{\prime}$ ($l=0, 1$).
Explicit expressions of the Landau parameters for the extended Skyrme
energy functional can be found in Ref.~\cite{Cha09}.

The Landau stability conditions
\begin{eqnarray}
F_l          & > & -(2l+1),\\
F_l^{\prime} & > & -(2l+1),\\
G_l          & > & -(2l+1),\\
G_l^{\prime} & > & -(2l+1),
\end{eqnarray}
guarantee the stability of symmetric nuclear matter against distortions of the momentum distribution
functions in different channels. It is of particular interest to see that the
conditions on $F_0$, $F_0^{\prime}$, $G_0$ and $G_0^{\prime}$ of symmetric nuclear matter
ensure the stabilities of symmetric nuclear matter against spinodal instability, isospin
instability, ferromagnetic instability and spin-isospin instability,
respectively, which can be easily seen through the following
relationships~\cite{Ben02,Gor10}
\begin{eqnarray}
\frac{\hbar^2k_F^2}{3m^{\ast}_s}(1+F_0)  &=& \left.\frac{\partial ^2 \mathcal{E}(\rho,\rho_1,s_0,s_1)}{\partial \rho^2}\right\vert_{\rho_1=s_0=s_1=0},\label{Eq:F0}\\
\frac{\hbar^2k_F^2}{3m^{\ast}_s}(1+F_0^{\prime})&=& \left.\frac{\partial ^2 \mathcal{E}(\rho,\rho_1,s_0,s_1)}{\partial (\rho_1)^2}\right\vert_{\rho_1=s_0=s_1=0},\label{Eq:F0P} \\
\frac{\hbar^2k_F^2}{3m^{\ast}_s}(1+G_0) &=& \left.\frac{\partial ^2
\mathcal{E}(\rho,\rho_1,s_0,s_1)}{\partial (s_0)^2}\right\vert_{\rho_1=s_0=s_1=0}, \\
\frac{\hbar^2k_F^2}{3m^{\ast}_s}(1+G_0^{\prime}) &=& \left.\frac{\partial ^2 \mathcal{E}(\rho,\rho_1,s_0,s_1)}{\partial (s_1)^2}\right\vert_{\rho_1=s_0=s_1=0},
\end{eqnarray}
where $\mathcal{E}$ represents the energy density of nuclear matter.
Here, in the notation of Ref.~\cite{Ben02}, $\rho_1=(\rho_n-\rho_p)$
is the isovector scalar density,
$s_0=(\rho_{n\uparrow}-\rho_{n\downarrow}+\rho_{p\uparrow}-\rho_{p\downarrow} )$
is the isoscalar vector density and
$s_1=(\rho_{n\uparrow}-\rho_{n\downarrow}-\rho_{p\uparrow}+\rho_{p\downarrow} )$
the isovector vector density with $\uparrow$ ($\downarrow$) denoting spin-up (-down).
For usual nuclear matter with spin saturation, one has the EOS
$E(\rho, \delta)=\mathcal{E}(\rho,\rho_1,0,0)/\rho$, the incompressibility
$K(\rho)=18\rho\frac{\partial E}{\partial\rho}+9\rho^2\frac{\partial ^2 E}{\partial \rho^2}$
and $E_{\mathrm{sym}}(\rho)=\frac{1}{2}\frac{\partial ^2 E}{\partial \delta}\vert_{\delta=0}$,
and then one can obtain from Eqs.~(\ref{Eq:F0}) and (\ref{Eq:F0P}) the following relations
\begin{eqnarray}
K(\rho) &=& \frac{3\hbar^2 k_{F}^2}{m_{s}^{\ast}}(1+F_0), \label{Eq:KLandau}\\
E_{\mathrm{sym}}(\rho)&=&\frac{\hbar ^2 k_{F}^2}{6m_{s}^{\ast}}(1+F_0^{\prime}).
\end{eqnarray}
In addition, $F_1$ and $F_1^{\prime}$ are directly related to
the isoscalar and isovector effective masses through
the following expressions~\cite{Gor10}
\begin{eqnarray}
\frac{m_{s}^{\ast}}{m}&=&1+\frac{F_1}{3}, \label{Eq:MsLandau} \\
\frac{m_s^{\ast}}{m_v^{\ast}}&=&1+\frac{F_1^{\prime}}{3}.
\end{eqnarray}
The conditions $F_1 >-3$ and $F_1^{\prime}>-3$ are thus naturally
satisfied at arbitrary densities for positive isoscalar and isovector
effective masses. Moreover, under the assumption $|E| \ll mc^2$, the
sound velocity $v_s$ in nuclear matter can be obtained from the
relation $mv_s^2\approx K(\rho)/9$, and thus in symmetric nuclear matter,
from Eqs.~(\ref{Eq:KLandau}) and (\ref{Eq:MsLandau}), one can also express
the $v_s$ in terms of Landau parameters as~\cite{Mar02}
\begin{equation}
mv_s^2\approx\frac{\hbar^2k_F^2}{3m}\frac{1+F_0}{1+\frac{1}{3}F_1}. \label{Eq:Sound}
\end{equation}

For pure neutron matter, there are only four Landau parameters $F_l^{(n)}$
and $G_l^{(n)}$ $(l=0, 1)$. Similarly the conditions $G_0^{(n)}>-1$ and
$G_1^{(n)}>-3$ guarantee the stability of pure neutron matter against spin
polarization --- or ferromagnetic transition. Explicit expressions for
Landau parameters of pure neutron matter can be found in Ref.~\cite{Gor10}.
Generally, a critical density $\rho_{cr}$ can be defined as the maximum
density below which all the twelve Landau parameters of symmetric nuclear
matter and pure neutron matter
(except $F_0$ for symmetric nuclear matter at subsaturation density
which leads to the well-known spinodal instability) satisfy the stability conditions.

For Skyrme interactions, as we pointed out before,
to remove the spin and spin-isospin instabilities, one can omit the time-odd terms
in $(\bm{s}_n+\bm{s}_p)\times(\bm{T}_n+\bm{T}_p)$ and
$(\bm{s}_n-\bm{s}_p)\times(\bm{T}_n-\bm{T}_p)$, namely set
$C_0^T=C_1^T=0$ in the notation of Ref.~\cite{Ben02}.
It should be noted that, as mentioned in Ref.~\cite{Cha10}, with this prescription, the
Landau parameters $G_1$, $G_1^{\prime}$ and $G_1^{(n)}$ all vanish,
leading to unrealistic effective masses in polarized matter.
In addition, if one sets $C_0^T=C_1^T=0$, the associated time-even terms in $J^2$
and $J_q^2$ also should be dropped for self-consistence.
In the present work, we will impose $\rho_{cr} > 1.2$ fm$^{-3}$ in the fitting
procedure to construct the new parameter sets of the extended Skyrme interactions.

\subsection{Isoscalar giant monopole resonance}

The energy of the isoscalar giant monopole resonance --- or the
breathing mode --- perhaps is the most important and efficient probe of
the incompressibility of nuclear matter around $\rho_0$. Thus we also
include in our following fit the experimental data of the ISGMR
energy for several spherical nuclei. The ISGMR energy is evaluated as
\begin{equation}
E_{\mathrm{GMR}}=\sqrt{\frac{m_1}{m_{-1}}}, \label{Eq:EGMR}
\end{equation}
where $m_i$ are $i$-th energy weighted sum rules defined as
\begin{equation}
m_i=\sum_{\nu}\vert \langle \nu\vert \hat{F} \vert 0 \rangle \vert^2 \left(E_{\nu} \right)^i.
\end{equation}
Here $\vert \nu \rangle$  is the RPA excitation state for the monopole operator $\hat{F}=\sum_i^{A} r_i^2$.

It is well established that the energy weighted sum rule $m_1$ can be
evaluated as~\cite{Bohr75}
\begin{equation}
m_1 =2\frac{\hbar^2}{m}A\langle r^2 \rangle, \label{Eq:m1}
\end{equation}
where A is the nucleon number, $m$ is the nucleon mass and $\langle r^2 \rangle$
is the ground-state rms radius. The moment $m_{-1}$ can be calculated through
the constrained-HF (CHF) approach~\cite{Boh79, Sil06}
\begin{equation}
m_{-1} = -\left.\frac{1}{2} \frac{d}{d\lambda}\langle\lambda \vert r^2 \vert \lambda \rangle ^2
\right\vert_{\lambda=0}, \label{Eq:m-1}
\end{equation}
where $\vert \lambda \rangle$ is the HF ground state for the CHF Hamiltonian
$\hat{H}+\lambda \hat{F}$. In this work, we calculate the ISGMR energy using
Eqs.~(\ref{Eq:EGMR}), (\ref{Eq:m1}) and (\ref{Eq:m-1}).

\section{Fitting Strategy}
\label{Sec:Fit}

In the present work, we use the simulated annealing method ~\cite{Agr05}
to determine the parameters of the extended Skyrme interactions by
minimizing the weighted sum of squared errors
\begin{equation}
\chi^2 =\sum_{i=1}^{N_d}\left( \frac{M_i^{\mathrm{exp}}-M_i^{\mathrm{th}}}{\sigma_i}\right)^2,
\end{equation}
where $N_d$ is the number of experimental data points, $M_i^{\mathrm{exp}}$
and $M_i^{\mathrm{th}}$  are the experimental and theoretical values for a
selected observable, respectively, and $\sigma_i$ is the adopted error
which is used to balance the relative weights of the various types
of observables.

We include the following experimental data of a number of spherical
even-even nuclei in the fit:
(i) the binding energies $E_B$ of $12$ nuclei, namely, $^{16}$O, $^{40,48}$Ca, $^{56, 68}$Ni, $^{88}$Sr,
$^{90}$Zr, $^{100,116, 132}$Sn, $^{144}$Sm, $^{208}$Pb~\cite{Wan12};
(ii) the charge rms radii $r_c$ of  $10$ nuclei, namely, $^{16}$O, $^{40,48}$Ca, $^{56}$Ni, $^{88}$Sr,
$^{90}$Zr, $^{116, 132}$Sn, $^{144}$Sm, $^{208}$Pb~\cite{Ang04, Bla05};
(iii) the ISGMR energies $E_{\mathrm{GMR}}$ of $^{90}$Zr, $^{116}$Sn, $^{144}$Sm
and $^{208}$Pb~\cite{You99}.
To regulate
the respective $\chi^2$ for each sort of observable to be roughly equal
to the number of corresponding data points, we assign a theoretical
error $1.5$ MeV to $E_B$, $0.015$ fm to $r_c$ while use the experimental
error multiplied by a factor $3.54$ for ISGMR energy $E_{\mathrm{GMR}}$.
In addition, the following constraints are considered in the optimization:
(i) the critical density $\rho_{cr}$ should be greater than $1.2$ fm$^{-3}$;
(ii) the neutron $3p_{1/2}-3p_{3/2}$ energy level splitting in $^{208}$Pb
should lie in the range of $0.8-1.0$ MeV;
(iii) the pressure of symmetric nuclear matter should be consistent with
the constraints in the density region of $2\rho_0 <\rho<4.6\rho_0$ obtained
from analyzing flow data in heavy-ion collisions~\cite{Dan02};
(iv) the EOS of pure neutron matter should conform to the predictions of the
latest chiral effective field theory calculations with controlled
uncertainties~\cite{Tew13}.
Furthermore, we fix the values of the magnitude $E_{\mathrm{sym}}(\rho_c)$
and density slope $L(\rho_c)$ of the symmetry energy at $\rho_c=0.11$ fm$^{-3}$
to be equal to those extracted from the isotope binding energy
difference~\cite{Zha13} and the electric dipole polarizability in
$^{208}$Pb~\cite{Zha14}, i.e. $E_{\mathrm{sym}}(\rho_c) = 26.65$ MeV and
$L(\rho_c)=47.3$ MeV.
For isoscalar and isovector effective masses at saturation density,
$m_{s,0}^{\ast}$ and $m_{v,0}^{\ast}$, we consider three different cases:
(i) $m_{s,0}^{\ast}=0.9m$ and $m_{v,0}^{\ast}=0.75m$ in parameter set eMSL09,
which conform to the constraints we extracted recently from analyzing the
giant resonances in $^{208}$Pb~\cite{Zha15a};
(ii) $m_{s,0}^{\ast}=0.8m$ and $m_{v,0}^{\ast}=0.7m$ in parameter set eMSL08;
(iii) $m_{s,0}^{\ast}=0.7m$ and $m_{v,0}^{\ast}=0.6m$ in parameter set eMSL07.
For eMSL08 and eMSL07, the condition $m_{s,0}^{\ast}-m_{v,0}^{\ast}=0.1m$
is imposed to be consistent with the extraction from global nucleon optical
potentials constrained by world data on nucleon-nucleus and (p,n)
charge-exchange reactions~\cite{Xu10,Li15a}.

\section{Results and discussions}
\label{Sec:Result}

Using the fitting procedure described in the previous section,
we obtain three extended Skyrme interactions, namely, eMSL07,
eMSL08 and eMSL09. Tab.~\ref{Tab:Skyrme} lists the values of the
Skyrme parameters and the corresponding $\chi^2$.
In the following, we discuss their performances in describing
properties of spherical nuclei, nuclear matter and neutron stars.
For comparison, we also show
the corresponding results obtained with three typical
interactions from Brussels group~\cite{Gor13}, i.e., BSk22, BSk24 and BSk26,
which adopt the similar extended Skyrme interaction as in our present work and
have been used to construct the
HFB mass models. The BSk22  and BSk24 parameter sets are adjusted
to fit the EOS of neutron matter obtained by BHF calculations
using Argonne V$_{18}$ two-body force with
compatible microscopic nuclear three-body forces~\cite{Li08},
while BSk26 is adjusted to fit the well known
Akmal-Pandharipande-Ravenhall (APR) EOS of neutron matter~\cite{Akm98}.
The symmetry energy value at saturation density is fixed at $E_{\text{sym}}(\rho_0)=32$ MeV
for BSk22 and at $E_{\text{sym}}(\rho_0)=30$ MeV for both BSk24 and BSk26.
In addition, the BSk family are also constrained to reproduce several
properties of nuclear matter and to support the heaviest observed
neutron stars~\cite{Gor13}. For properties of finite nuclei,
the BSk family are constructed based on Hartree-Fock-Bogoliubov (HFB)
calculations with a microscopic pairing force and includes
several additional corrections (e.g., the phenomenological Wigner terms
and correction terms for the spurious collective energy)
to better fit nuclear masses.

\begin{table}[bp]
\caption{Skyrme parameters and $\chi^2$ of the extended Skyrme
parameter sets eMSL07, eMSL08 and eMSL09: lines 1--4 show
the $\chi^2$ evaluated from experimental data of binding energies,
charge radii and ISGMR energies, namely $\chi ^2_{E_B}$,
$\chi ^2_{r_c}$ and $\chi ^2_{E_{\mathrm{GMR}}}$, together
with the total $\chi^2$, $\chi^2_{\mathrm{tot}} $;
lines 5--18 show the Skryme parameters. The last two lines
show the calculated neutron skin thickness $\Delta r_{np}^{208}$
 and neutron $3p_{1/2}-3p_{3/2}$ energy level
splitting $\epsilon_{ls}^{208}$ of $^{208}$Pb for the three interactions. }
\label{Tab:Skyrme}
\begin{tabular}{lccc}
\hline \hline
 &eMSL07 	 &eMSL08 & eMSL09		\\
\hline
$\chi^2_{\mathrm{tot}} $&24.48&23.45&22.68 \\
$\chi ^2_{E_B}$&13.47 &10.07 &8.25    \\
$\chi ^2_{r_c}$& 7.16  &9.17 & 10.17   \\
$\chi ^2_{E_{\mathrm{GMR}}}$&3.85&4.22 & 4.26 \\
$t_0(\mathrm{MeV}\cdot \mathrm{fm}^{3})$&              -2941.76     & -2429.09   & -2231.73	\\
$t_1(\mathrm{MeV}\cdot \mathrm{fm}^{5})$&               575.338     &  493.720   &  431.073	\\
$t_2(\mathrm{MeV}\cdot \mathrm{fm}^{5})$&              -398.554     & -424.380   & -426.692	\\
$t_3(\mathrm{MeV}\cdot \mathrm{fm}^{3+3\alpha})$&       16403.9     &  14502.7   &  14248.9  \\
$t_4(\mathrm{MeV}\cdot \mathrm{fm}^{5+3\beta})$&       -773.045     & -724.215   & -703.052  \\
$t_5(\mathrm{MeV}\cdot \mathrm{fm}^{5+3\gamma})$&       1159.83     &  887.873   &  668.913  \\
$x_0$                                           &      0.371259     & 0.334701   & 0.280754	\\
$x_1$&                                                 0.137412     & 0.132739   & 0.352663	\\
$x_2$&                                                -0.713811     &-0.666437   &-0.603887	\\
$x_3$&                                                 0.412960     & 0.337131   & 0.177830	\\
$x_4$&                                                 0.0852272    & 0.0754104  & 0.180194	\\
$x_5$&                                                -0.714565     &-0.579724   &-0.361533	\\
$\alpha$&                                              0.132019     & 0.191960   & 0.231193	\\
$\beta$&                                                1           &  1		    &  1		\\
$\gamma$&                                               1           &  1         &  1		\\
$W_0(\mathrm{MeV}\cdot \mathrm{fm}^{5})$&               118.15     &  110.85	& 101.53	\\
$\Delta r_{np}^{208}$ (fm)                           & 0.182       & 0.183      & 0.183     \\
$\epsilon_{ls}^{208}$(MeV)                           & 0.99        & 0.90       & 0.81      \\
\hline\hline
\end{tabular}
\end{table}

\begin{figure}[bp]
\includegraphics[width=1.0\linewidth]{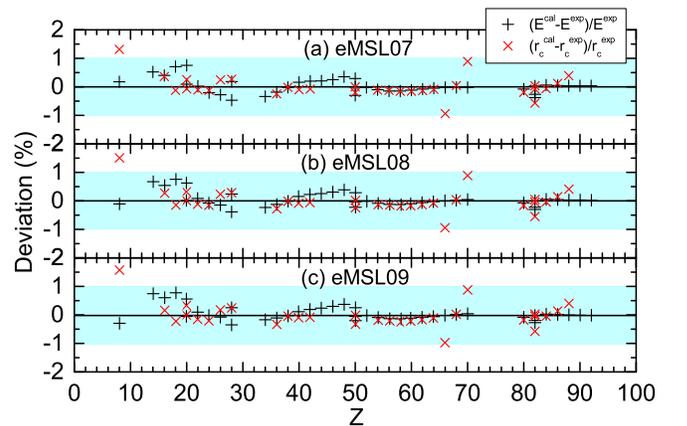}
\caption{(Color online) Deviations of the binding energies (plus symbols) and charge rms radii
(cross symbols) of a number of nuclei with atomic number $Z$ obtained from SHF
calculations with eMSL07, eMSL08 and eMSL09 from those measured
in experiments~\cite{Wan12,Ang04,Bla05}. The bands indicate a deviation within $\pm1\%$. }
\label{Fig:EbRc}
\end{figure}

\subsection{Properties of finite nuclei}

In Fig.~\ref{Fig:EbRc}, we present the relative deviations of
the binding energies and charge rms radii for
a number of spherical nuclei calculated with the
three new interactions eMSL07, eMSL08 and eMSL09 from the
corresponding experimental data~\cite{Wan12,Ang04,Bla05}.
For the binding energy, we show the results for $41$ spherical
even-even nuclei for which the data are available,
i.e.,
$_{~8}^{16}{\mathrm{O}}$, $_{14}^{34}{\mathrm{Si}}$,
$_{16}^{36}{\mathrm{S}}$, $_{18}^{38}{\mathrm{Ar}}$,
$\ce{^{40,48}_{20}Ca}$, $\ce{^{50}_{22}Ti}$,
$\ce{^{52}_{24}Cr}$, $\ce{^{54}_{26}Fe}$,
$\ce{^{56,68}_{28}Ni}$, $\ce{^{84}_{34}Se}$,
$\ce{^{86}_{36}Kr}$, $\ce{^{88}_{38}Sr}$,
$\ce{^{90}_{40}Zr}$, $\ce{^{92}_{42}Mo}$,
$\ce{^{94}_{44}Ru}$, $\ce{^{96}_{46}Pd}$,
$\ce{^{98}_{48}Cd}$, $\ce{^{100,116,132}_{50}Sn}$,
$\ce{^{134}_{52}Te}$, $\ce{^{136}_{54}Xe}$,
$\ce{^{138}_{56}Ba}$, $\ce{^{140}_{58}Ce}$,
$\ce{^{142}_{60}Nd}$, $\ce{^{144}_{62}Sm}$,
$\ce{^{146}_{64}Gd}$, $\ce{^{148}_{66}Dy}$,
$\ce{^{150}_{68}Er}$, $\ce{^{152}_{70}Yb}$,
$\ce{^{206}_{80}Hg}$, $\ce{^{198, 208, 214}_{82}Pb}$,
$\ce{^{210}_{84}Po}$, $\ce{^{212}_{86}Rn}$,
$\ce{^{214}_{88}Ra}$, $\ce{^{216}_{90}Th}$ and
$\ce{^{218}_{92}Pu}$;
for the charge rms radii, we show the results for
$31$ spherical even-even nuclei for which the data
are available, i.e.,
$_{~8}^{16}{\mathrm{O}}$, $_{16}^{36}{\mathrm{S}}$,
$_{18}^{38}{\mathrm{Ar}}$, $\ce{^{40,48}_{20}Ca}$,
$\ce{^{50}_{22}Ti}$, $\ce{^{52}_{24}Cr}$,
$\ce{^{54}_{26}Fe}$, $\ce{^{56}_{28}Ni}$,
$\ce{^{86}_{36}Kr}$, $\ce{^{88}_{38}Sr}$,
$\ce{^{90}_{40}Zr}$, $\ce{^{92}_{42}Mo}$,
$\ce{^{116,132}_{50}Sn}$, $\ce{^{136}_{54}Xe}$,
$\ce{^{138}_{56}Ba}$, $\ce{^{140}_{58}Ce}$,
$\ce{^{142}_{60}Nd}$, $\ce{^{144}_{62}Sm}$,
$\ce{^{146}_{64}Gd}$, $\ce{^{148}_{66}Dy}$,
$\ce{^{150}_{68}Er}$, $\ce{^{152}_{70}Yb}$,
$\ce{^{206}_{80}Hg}$, $\ce{^{198,208,214}_{82}Pb}$,
$\ce{^{210}_{84}Po}$, $\ce{^{212}_{86}Rn}$ and
$\ce{^{214}_{88}Ra}$.
The hatched bands indicate a deviation within $\pm 1\%$.
One can see that all the three new interactions can describe
reasonably (within $\pm 1\%$)
the ground-state properties of these spherical nuclei except for the light nucleus
$^{16}$O for which the deviations of charge rms radius
are about $1.5\%$.

It should be noted that in the fitting, we use only $12$ nuclei for
the binding energy data and $10$ nuclei for the charge rms radius data.
The standard deviation for the binding energy of the fitted $12$ nuclei
is $1.59$ MeV for eMSL07, $1.37$ MeV for eMSL08, $1.24$ MeV for eMSL09,
and it becomes $1.043$ MeV, $0.79$ MeV and $1.018$ MeV for BSk22,
BSk24 and BSk26~\cite{HFBmass}, respectively.
The standard deviation for the binding energy changes to
$1.81$ MeV for eMSL07, $1.63$ MeV for eMSL08, $1.49$ MeV for eMSL09,
$0.66$ MeV for BSk22, $0.56$ MeV for BSk24, and $0.66$ MeV for BSk26,
for the $41$ spherical even-even nuclei shown in Fig.~\ref{Fig:EbRc}.
For the charge rms radius, the standard deviation of the fitted $10$ nuclei
is $0.0127$ fm for eMSL07, $0.0144$ fm for eMSL08, $0.0151$ fm for eMSL09,
while it becomes $0.0292$ fm, $0.0255$ fm and $0.0333$ fm for BSk22,
BSk24 and BSk26~\cite{HFBmass}, respectively.
The standard deviation for the charge rms radius changes to
$0.0162$ fm for eMSL07, $0.0166$ fm for eMSL08, $0.0169$ fm for eMSL09,
$0.0245$ fm for BSk22, $0.0231$ fm for BSk24, and $0.0290$ fm for BSk26
for the $31$ spherical even-even nuclei shown in Fig.~\ref{Fig:EbRc}.
These results clearly indicate that
the BSk family make a significant improvement on the description of the binding
energy compared with the eMSL family, while the eMSL family give a better
description of the charge rms radius, at least for the spherical even-even
nuclei shown in Fig.~\ref{Fig:EbRc}.

We would also like to point out that, in our fits and calculations, the pairing effects
are treated within the framework of BCS theory with the constant gap approximation.
This treatment of the pairing should be reasonable for the fitted nuclei in the
present work since they are essentially (semi-)doubly magic nuclei for which
the pairing effects are unimportant. For the nuclei with neutron or proton number
deviating from magic number, the pairing effects might be important, and the more sophisticated Bogoliubov treatment with a microscopic
pairing force as well as some additional corrections, such as the Wigner terms
and the spurious collective energy, can better
reproduce the binding energy over the whole nuclear
chart as demonstrated in the HFB calculations (see e.g. Ref.~\cite{Cha09,Gor10,Gor13,Gor15} and
references therein).
Therefore, while the present eMSL family can reasonably describe
the binding energies and charge rms radii of the ground-state spherical even-even nuclei,
using the more sophisticated treatment of the pairing effects and including some additional
corrections (e.g., the Wigner terms and the spurious collective energy) are important to
improve the description of the binding energy over the whole nuclear chart. In addition,
in the present work, we only focus on the spherical nuclei without considering nuclear
deformation which should be important for the description of nuclei in the whole nuclear chart.
Nevertheless, a very accurate
description of the binding energy over the whole nuclear chart is beyond the scope of the
present work and this could be pursued in future.

Furthermore, the first three rows in Tab.~\ref{Tab:Skyrme}
 show that the variation of the effective masses has little impact
on the fitting quality for the data of the binding energies and charge
radii. Moreover, the $\chi^2$ for the binding energy and charge radius
calculated with the three interactions (see Table~\ref{Tab:Skyrme})
suggest that the binding energy data prefer a larger $m_{s,0}^{\ast}$
while the charge radius data favor a smaller $m_{s,0}^{\ast}$,
and these features are consistent with the results in the SHF
model with the standard Skyrme interactions~\cite{Klu09}.

The neutron skin thickness
$\Delta r_{np}=\langle r_n^2 \rangle ^{1/2}- \langle r_p^2 \rangle ^{1/2}$,
i.e., the difference of the neutron and proton rms radii, is of particular
importance for the study of the density dependence of the symmetry energy.
Mean field calculations using many different relativistic and nonrelativistic
interactions have indicated that the neutron skin thickness is strongly
correlated to the density slope $L(\rho_0)$ of the symmetry energy at
saturations density~\cite{Bro00,Typ01,Che05}. Moreover, in Ref.~\cite{Zha13},
we find the neutron skin thickness of heavy nuclei is uniquely determined
by $L(\rho_c)$ at a subsaturation density $\rho_c=0.11$ fm$^{-3}$.
So we show in row 19 of Tab.~\ref{Tab:Skyrme} the
neutron skin thickness of $^{208}$Pb, $\Delta r_{np}^{208}$, predicted by
HF calculations using the three new interactions. Due to the imposed
condition $L(\rho_c)=47.3$ MeV, the eMSL family predict quite similar
values of $\Delta r_{np}^{208}$, i.e., about $0.18$~fm, which are in very
good agreement with the $\Delta r_{np}^{208}=0.170\pm0.016$~fm and $\Delta r_{np}^{208}
=0.176\pm0.027$~fm obtained, respectively, from SHF analyses of neutron skin data of Sn isotopes~\cite{Zha13}
and experimental data of the electric dipole polarizability
$\alpha_{\mathrm{D}}$ in $^{208}$Pb~\cite{Zha14}.
These results are also consistent with the estimated range
$\Delta r_{np}^{208}=0.165\pm(0.009)_{\mathrm{expt}}\pm(0.013)_{\mathrm{theor}}
\pm(0.021)_{\mathrm{est}}$ fm extracted from the measured $\alpha_{\mathrm{D}}$
in $^{208}$Pb~\cite{Roc13}, the constraint $\Delta r_{np}^{208}=
0.15\pm0.03(\mathrm{stat.})^{+0.01}_{-0.03}(\mathrm{sys.})$ fm
from coherent pion photoproduction cross sections\cite{Tar14}, and the
constraint $\Delta r_{np}^{208}=0.33^{+0.16}_{-0.18}$ fm extracted
from PREX at JLab~\cite{Abr12}.
We note that the BSk22 predicts a neutron skin thickness of
$^{208}$Pb, i.e., $\Delta r_{np}^{208} =0.18$~fm~\cite{Gor13},
while both BSk24 and BSk26 predict a little smaller value of
$\Delta r_{np}^{208} =0.14$~fm~\cite{Gor13}.
This feature can be
understood from the fact that the $\Delta r_{np}^{208}$ is uniquely
determined by $L(\rho_c)$ at a subsaturation density
$\rho_c=0.11$ fm$^{-3}$~\cite{Zha13}, and both BSk24 and BSk26 have a
smaller value of $L(\rho_c)=38.1$ MeV while BSk22 has a value of
$L(\rho_c)=49.3$ MeV (see Table~\ref{Tab:MSL} in the following).

\begin{table}[bp]
\caption{Comparison of the ISGMR energies $E_{\mathrm{GMR}}=
\sqrt{m_1/m_{-1}}$ (in MeV) in $^{90}$Zr, $^{116}$Sn, $^{144}$Sm
and $^{208}$Pb obtained for the eMSL07, eMSL08 and eMSL09
interactions with the experimental data~\cite{You99, Li07, Pat13}.}
\label{Tab:ISGMR}
\begin{tabular}{lccccc}
\hline
\hline
Nucleus    & TAMU & RCNP & eMSL07 & eMSL08 & eMSL09 \\
\hline
$^{90}$Zr  & $17.81\pm0.35$ &---& 17.65 & 17.68 & 17.70 \\
$^{116}$Sn & $15.90\pm0.07$ &$15.70\pm0.10$& 16.19 & 16.20 & 16.21 \\
$^{144}$Sm & $15.25\pm0.11$ & ---& 15.28 & 15.27 & 15.29 \\
$^{208}$Pb & $14.18\pm0.11$ &$13.50\pm0.10$& 13.57 & 13.53 & 13.54 \\
\hline
\hline
\end{tabular}
\end{table}

The single-particle spectra are also important observables and have profound
impacts on the properties of super-heavy nuclei~\cite{Ben01}.
However, it is known that the HF approach can not well describe the single-particle
energies due to the self-interaction error~\cite{Ben03}, and thus we
have not included them in our present fit. Unlike the single-particle energies,
their differences among particle states or hole states (e.g.,
the spin-orbit splitting without crossing the shell gap), are believed to be
robust observables which can be safely compared with the results of HF
calculations~\cite{Ben03}. Here we list in last row of Tab.~\ref{Tab:Skyrme}
the neutron $3p_{1/2}-3p_{3/2}$ energy level splitting in $^{208}$Pb,
$\epsilon_{ls}^{208}$, for the eMSL family. As one can see, the results,
especially $\epsilon_{ls}^{208}=0.90$~MeV for eMSL08, well agree with the
experimental value $0.89$~MeV~\cite{Vau72}.

For the ISGMR energy, we show in Tab.~\ref{Tab:ISGMR} the
calculated results together with the corresponding experimental data
reported by TAMU group~\cite{You99} and RCNP group~\cite{Li07, Pat13}.
It can be seen that the three new interactions predict very similar
and overall reasonable monopole response properties.

\subsection{Properties of nuclear matter}

\begin{figure*}[!]
\includegraphics[width=0.95\linewidth]{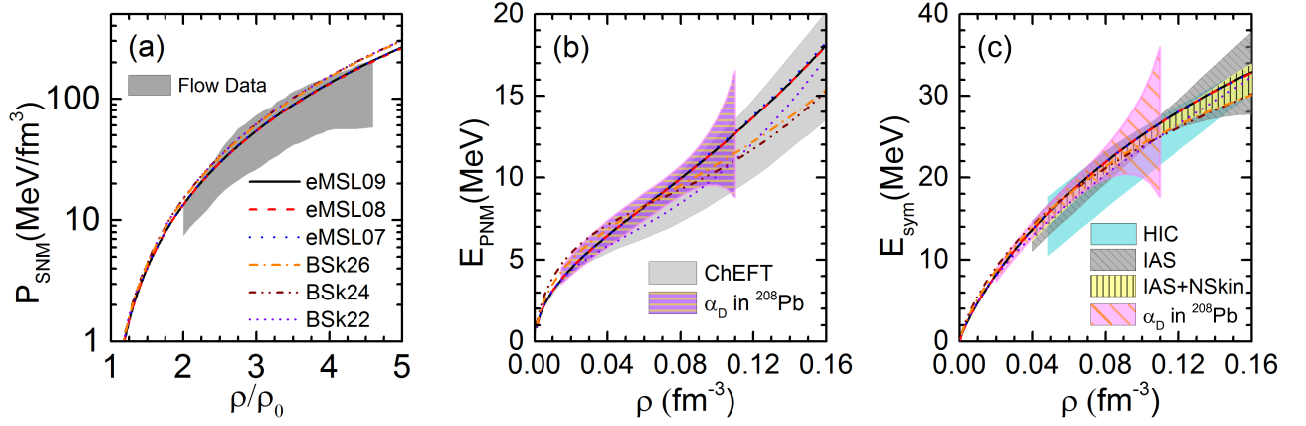}
\caption{(Color online) Pressure of symmetric nuclear matter
$P_{\mathrm{SNM}}(\rho)$ (a), the binding energy per nucleon of pure
neutron matter $E_{\mathrm{PNM}}(\rho)$ (b) and the symmetry energy
$E_{\mathrm{sym}}(\rho)$ (c) as functions of densities obtained from
HF calculations using eMSL07, eMSL08, eMSL09, BSk22, BSk24 and BSk26.
Several constraints from analyzing experimental data and the predictions
of theoretical calculations are also included for comparison
(see text for the details).}
\label{Fig:EOS}
\end{figure*}

In this subsection we discuss the properties of infinite nuclear matter
predicted by the three new extended Skyrme interactions. We present in
Tab.~\ref{Tab:MSL} the values of a number of macroscopic quantities in
the SHF model with MSL07, eMSL08 and eMSL09.
The corresponding results with the BSk22, BSk24
and BSk26 interactions~\cite{Gor13}
are also included in Tab.~\ref{Tab:MSL} for comparison.
One can
see that the eMSL family predict very similar
and reasonable macroscopic quantities at saturation density.
In particular, the values of the magnitude and density slope of the
symmetry energy, $E_{\mathrm{sym}}(\rho_0)$ and $L(\rho_0)$, are essentially
consistent with other constraints obtained from analyzing terrestrial
experiments and astrophysical observations as well as the predictions
of theoretical calculations~\cite{Tsa12,Lat12,ChenLW12,LiBA12,Tew13}.
The calculated results of higher-order coefficients $J_0$ and $K_{\mathrm{sym}}(\rho_0)$ with the eMSL family
are also in very good agreement with their empirical values~\cite{Che09,ChenLW11,Cai14}.
Compared with the eMSL family, the BSk family give larger $K_0$ and $J_0$, and thus predict
stiffer EOSs of SNM, especially at high densities, which may have considerable
effects on neutron star properties. Moreover, the BSk family predict rather different
values of isovector macroscopic quantities $L(\rho_0)$ and $K_{\mathrm{sym}}(\rho_0)$
which suggests they lead to very different density dependence of the symmetry energy
especially in high-density region. In the following, we will give a more detailed discussion
on the EOSs obtained with the six extended Skyrme interactions.

In the last row of Tab.~\ref{Tab:MSL}, we also show the values
of the second-order symmetry coefficient $K_{\mathrm{sat,2}} = K_{\text {sym}} - (6+J_0/K_0)L$ for the
isobaric incompressibility of asymmetric nuclear matter~\cite{Che09}, for the six
extended Skyrme interactions. It is generally believed that the coefficient
$K_{\mathrm{sat,2}}$ can be extracted from the ISGMR energy $E_{\mathrm{GMR}}$
of neutron-rich nuclei which is related to the incompressibility
of finite nuclei, $K_A$, through the relation~\cite{Bla80}
\begin{equation}
E_{\mathrm{GMR}} = \sqrt{\frac{\hbar ^2}{m\langle r^2 \rangle}K_A}.
\end{equation}
In the leptodermous approximation, $K_A$ can be expressed as~\cite{Bla80}
\begin{equation}
\label{Eq:KA}
K_A=K_{\mathrm{v}}+K_{\mathrm{s}}A^{-1/3}+K_{\tau}\left( \frac{N-Z}{A}\right)^2 +K_{\mathrm{c}}\frac{Z^2}{A^{4/3}}
\end{equation}
with $K_{\mathrm{v}}$, $K_{\mathrm{s}}$, $K_{\tau}$, $K_{\mathrm{c}}$
being the volume, surface, isospin and Coulomb terms, respectively.
Here high order terms in powers of $A^{1/3}$ and $(N-Z)/A$ are neglected,
and the $K_{\tau}$ parameter is usually thought
to be equivalent to the $K_{\mathrm{sat,2}}$ parameter. By fitting
ISGMR energies of Sn and Cd isotopes using the above formula,
Li \textit{et al.} and Patel \textit{et al.} obtained
$K_{\tau}=-550\pm100$ MeV and $K_{\tau}=-555\pm75$ MeV,
respectively~\cite{Li07,Li10,Pat12}. And a recent analysis of the
ISGMR experimental data by Stone \textit{et al.} leads to
a constraint of $-840 < K_{\tau} <-350$~MeV with a large uncertainty~\cite{Sto14}.
The magnitudes of these constrained $K_{\tau}$ are essentially larger
than those of $K_{\mathrm{sat,2}}$ predicted by the three new
interactions as well as the BSk interactions.
At this point, it should be pointed out that, since the extracted
values of $K_{\tau}$ are contaminated by contributions of high-order
terms which are neglected in Eq.~(\ref{Eq:KA}), one should not directly
compare them against the value of $K_{\mathrm{sat,2}}$ from the SHF
calculations~\cite{Pea10,Sto14}.
From Tab.~\ref{Tab:MSL}, one can see that while the eMSL family predict
similar $K_{\mathrm{sat,2}}$, the BSk family, especially BSk24, predict
larger values of $K_{\mathrm{sat,2}}$.
It is interesting to see that the
predictions of both eMSL family and BSk family are all in
very good agreement with the estimated value of $K_{\mathrm{sat,2}}=-370\pm120$~MeV
obtained from a standard SHF model~\cite{Che09} analysis on the
symmetry energy constrained in heavy-ion collisions~\cite{ChenLW05,Tsa09}.
To date the accurate determination of $K_{\mathrm{sat,2}}$
remains an open challenge.

\begin{table*}[!tbp]
\caption{Macroscopic quantities in the SHF model with eMSL07, eMSL08 and eMSL09.
The corresponding results from BSk22, BSk24, and BSk26~\cite{Gor13} are also included for comparison}.
\label{Tab:MSL}
\begin{tabular}{lcccccc}
\hline \hline
										 &  eMSL07 	&  eMSL08 &  eMSL09	&  BSk22   &  BSk24   &  BSk26 \\
\hline																			
$\rho_0(\mathrm{fm}^{-3}) $              &  0.1584  &  0.1584 &  0.1583 &  0.1578  &  0.1578  &  0.1589 \\
$E_0(\rho_0)(\mathrm{MeV})$              & -16.043  & -16.040 & -16.034 & -16.088  & -16.048  & -16.064  \\
$K_0(\mathrm{MeV})$                      &  229.7   &  229.0  &  229.6  &  245.9   &  245.5   &  240.8  \\
$J_0(\mathrm{MeV})$                      & -339.9   & -351.4  & -352.7  & -275.5   & -274.5   & -282.9  \\
$E_{\mathrm{sym}}(\rho_c)(\mathrm{MeV})$ &  26.65   &  26.65  &  26.65  &  25.03   &  24.98   &  25.3  \\
$L(\rho_c)(\mathrm{MeV})$                &  47.3    &  47.3   &  47.3   &  49.3    &  38.1    &  38.1   \\
$K_{\mathrm{sym}}(\rho_c)$(MeV)          & -94.7    & -92.7   & -91.1   & -32.5    & -67.3    & -103.5   \\
$G_S$(MeV$\cdot $fm$^{5}$)               &  118.8   &  104.2  &  92.8   &  116.6   &  113.4   &  118.6  \\
$G_V$(MeV$\cdot $fm$^{5}$)               &  58.2    &  49.9   &  58.4   &  21.2	   &  20.2	  &  38.2	 	\\
$G_{SV}$(MeV$\cdot $fm$^{5}$)			 & -9.0     & -8.3    & -9.5    & -6.2	   & -6.2	  &  3.83 	\\
$G_0^{\prime}(\rho_0)$					 & 	0.24    &  0.19   &  0.17   &  0.39	   &  0.40	  &  0.41	 	\\
$m_{s,0}^{\ast}/m$						 &  0.7     &  0.8    &  0.9    &  0.80	   &  0.80	  &  0.80	 	\\
$m_{v,0}^{\ast}/m$						 &  0.6     &  0.7    &  0.75   &  0.71	   &  0.71	  &  0.65	 	\\
$E_{\mathrm{sym}}(\rho_0)(\mathrm{MeV})$ &  32.7    &  32.8   &  32.8   &  32.0	   &  30.0	  &  30.0	 	\\
$L(\rho_0)(\mathrm{MeV})$                &  52.1    &  53.0   &  53.9   &  68.5	   &  46.4	  &  37.5	 	\\
$K_{\mathrm{sym}}(\rho_0)$(MeV)			 & -125.1   & -111.9  & -96.7   &  13.0	   & -37.6	  & -135.6	 	\\
$K_{\mathrm{sat,2}}$(MeV)			     & -360.4   & -348.8  & -337.2  & -321.2   & -264.0   & -316.6 \\
\hline
\hline
\end{tabular}
\end{table*}

To see more clearly the nuclear matter properties, we show in Fig.~\ref{Fig:EOS}
the pressure of symmetric nuclear matter $P_{\mathrm{SNM}}(\rho)$, the binding
energy per neutron in pure neutron matter $E_{\mathrm{PNM}}(\rho)$ and the symmetry
energy $E_{\mathrm{sym}}(\rho)$ as functions of densities for eMSL07, eMSL08,
eMSL09, BSk22, BSk24 and BSk26.
For comparison, Fig.~\ref{Fig:EOS} also includes the constraints on
$P_{\mathrm{SNM}}(\rho)$ in the density region of $2\rho_0$-4.6$\rho_0$ from analyzing
the flow data in relativistic heavy-ion collisions (Flow Data)~\cite{Dan02}; the predictions on
$E_{\mathrm{PNM}}(\rho)$ at subsaturation densities from ChEFT calculations
using next-to-next-to-next-to-leading order (N3LO) potential (ChEFT)~\cite{Tew13};
the constraints on the density dependence of the symmetry energy at subsaturation
densities from transport model analyses of mid-peripheral heavy-ion collisions of
Sn isotopes (HIC)~\cite{Tsa09} and the SHF analyses of isobaric analog states (IAS)
as well as combing additionally the neutron skin data (IAS+NSkin)~\cite{Dan14}.
In addition, the constraints ($\alpha_{\mathrm{D}}$ in $^{208}$Pb) on
$E_{\mathrm{PNM}}(\rho)$ and $E_{\mathrm{sym}}(\rho)$ extracted recently from analyzing
the electric dipole polarizability $\alpha_{\mathrm{D}}$ in $^{208}$ Pb~\cite{Zha15}
are also displayed in Fig.~\ref{Fig:EOS} (b) and (c), respectively.
It can be seen that,
compared with the BSk interactions,
the eMSL family predict relatively softer EOSs of SNM at supra-saturation
densities while stiffer symmetry energies at subsaturation densities.
We also notice that
the eMSL family predict very similar symmetric nuclear matter properties even up to
$5\rho_0$ as well as almost the same pure neutron matter EOS and the density dependence of the symmetry
energy at subsaturation densities. As we mentioned before,
the EOS of symmetric nuclear matter can be well constrained by the properties of finite nuclei and
experimental data of heavy-ion collisions. It is not surprising to see that the three new extended
Skyrme interactions predict almost the same $E_{\mathrm{PNM}}(\rho)$ and
$E_{\mathrm{sym}}(\rho)$ at subsaturation densities, as the magnitude and density
slope of the symmetry energy at $\rho_c=0.11$ fm$^{-3}$, namely
$E_{\mathrm{sym}}(\rho_c)$ and $L(\rho_c)$, are imposed to be $26.65$ MeV and $47.3$ MeV,
respectively. This condition guarantees the eMSL family predict reasonable
density dependence of the symmetry energy or the EOS of asymmetric nuclear matter
below and around $\rho_0$.

Overall, for the properties of nuclear matter below and around saturation density,
the eMSL and BSk family give quite similar predictions. The eMSL
family make a small improvement on the supra-saturation density behaviors of symmetric
nuclear matter for which the BSk interactions predict a little too large pressure around
$2\rho_0$ and above $4\rho_0$ compared with the constraint from flow data in
heavy-ion collisions~\cite{Dan02}, as shown in Fig.~\ref{Fig:EOS} (a). In addition,
the EOSs of pure neutron matter given by the eMSL family are
in better agreement with the recent constraint obtained from analyzing the
$\alpha_D$ data of $^{208}$Pb~\cite{Zha15} as well as the predictions from the
recent ChEFT calculations using N3LO
potential~\cite{Tew13}, as shown in Fig.~\ref{Fig:EOS} (b).

\begin{figure}[hbp]
~\\
\includegraphics[width=1\linewidth]{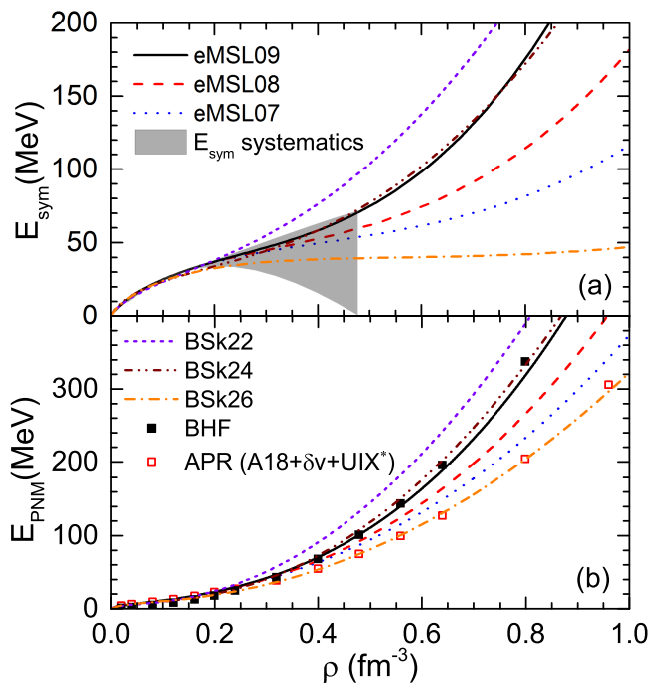}
\caption{(Color online) Density dependence of the symmetry energy
$E_{\mathrm{sym}}(\rho)$ (a) and  EOSs of pure neutron matter
$E_{\mathrm{PNM}}(\rho)$ (b) for the extended Skyrme
interactions eMSL07, eMSL08, eMSL09, BSk22, BSk24 and BSk26.
The shaded band in panel (a) is taken from Ref.~\cite{ChenLW15}.
The $E_{\mathrm{PNM}}(\rho)$ of APR is
taken from Ref.~\cite{Akm98} while that of BHF is from Ref.~\cite{Li08}.}
\label{Fig:Epnm}
\end{figure}

For high-density behaviors of the symmetry energy, we exhibit in the
upper panel of Fig.~\ref{Fig:Epnm} the symmetry energy $E_{\mathrm{sym}}(\rho)$
as a function of density up to $1$ fm$^{-3}$ for the eMSL and BSk family. One can see
that all these extended Skrme interactions except BSk26 predict a soft symmetry
energy below and around saturation density but a stiff symmetry energy at
supra-saturation densities. For the
BSk26 interaction, the symmetry energy
almost remains constant in the density region of $\rho>2\rho_0$.
In addition, the eMSL09 and BSk24 interactions predict very similar
high-density behaviors of the symmetry energy.
It is interesting to see
that the symmetry energy predictions from the eMSL family
are consistent with the constraint from very low density to $3\rho_0$ obtained recently by
extrapolating the well constrained subsaturation density symmetry energy based on
systematics of various relativistic and nonrelativistic EDFs~\cite{ChenLW15}.
Given the present poor
knowledge on the high-density behaviors of the symmetry
energy, we show in the lower panel of Fig.~\ref{Fig:Epnm} the comparison
of the pure neutron matter EOS $E_{\mathrm{PNM}}(\rho)$ as a function of density $\rho$
for the eMSL and BSk family with the predictions of nonrelativistic BHF calculations
using realistic Argonne V$_{18}$ force
together with compatible microscopic three microscopic
three-body forces~\cite{Li08} and the well known
APR EOS using the realistic A18+$\delta$v +
UIX$^{\ast}$ interaction~\cite{Akm98}. One can see that the extended Skyrme
interactions predict very reasonable neutron matter EOSs that are consistent
with the microscopic calculations.
In particular, the EOS of pure neutron matter with eMSL09 is
very close to that with BSk24 as well as the BHF calculations,
while those with eMSL07 and eMSL08 lie between the APR and BHF EOSs.

We would like to emphasize that the main difference of nuclear matter properties
for the three new extended Skyrme interactions is the high-density behaviors of
the EOS of asymmetric nuclear matter or the symmetry energy. This is partly
because of the variation of $m_{s,0}^{\ast}$ and $m_{v,0}^{\ast}$ values in these
extended Skyrme interactions, which has essential impacts
on the $\mathcal{H}_{\mathrm{eff}}$ (i.e., Eq.~(\ref{Eq:HEFF})) and is related
to the momentum dependence of the three body forces. Therefore, the additional
momentum and density dependent two-body forces which effectively simulate the
momentum dependence of the three-body force may play an important role for
the high-density behaviors of the EOS of asymmetric nuclear matter.
The three new extended Skyrme interactions constructed in the present work
predict very similar properties of nuclear matter below and around saturation
density but different high-density behaviors of the symmetry energy, and thus
are potentially useful for the investigation of the currently largely uncertain
high-density behaviors of the symmetry energy.

\subsection{Landau parameters}

\begin{figure}[hbp]
\includegraphics[width=1\linewidth]{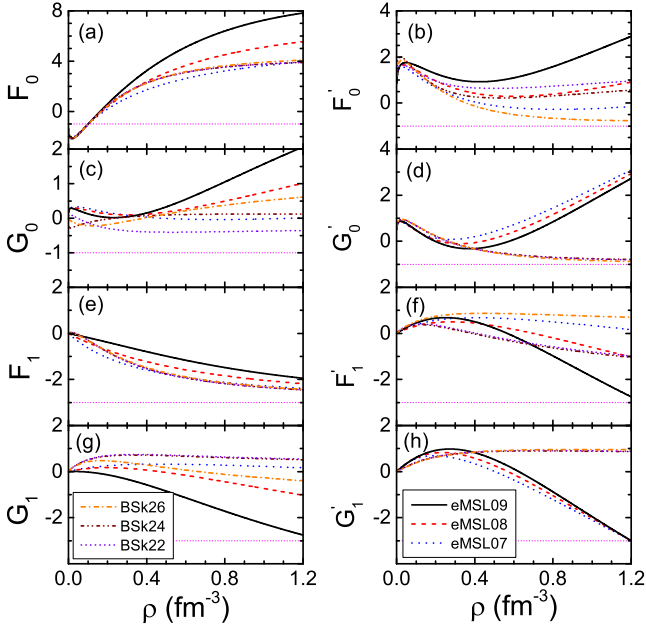}
\caption{(Color online) Density dependence of the Landau parameters
of symmetric nuclear matter for the extended Skyrme interactions
eMSL07, eMSL08, eMSL09, BSk22, BSk24 and BSk26. The short-dash-dotted
lines indicate the critical stability conditions.}
\label{Fig:LPsnm}
\end{figure}

\begin{figure}[h]
\includegraphics[width=1\linewidth]{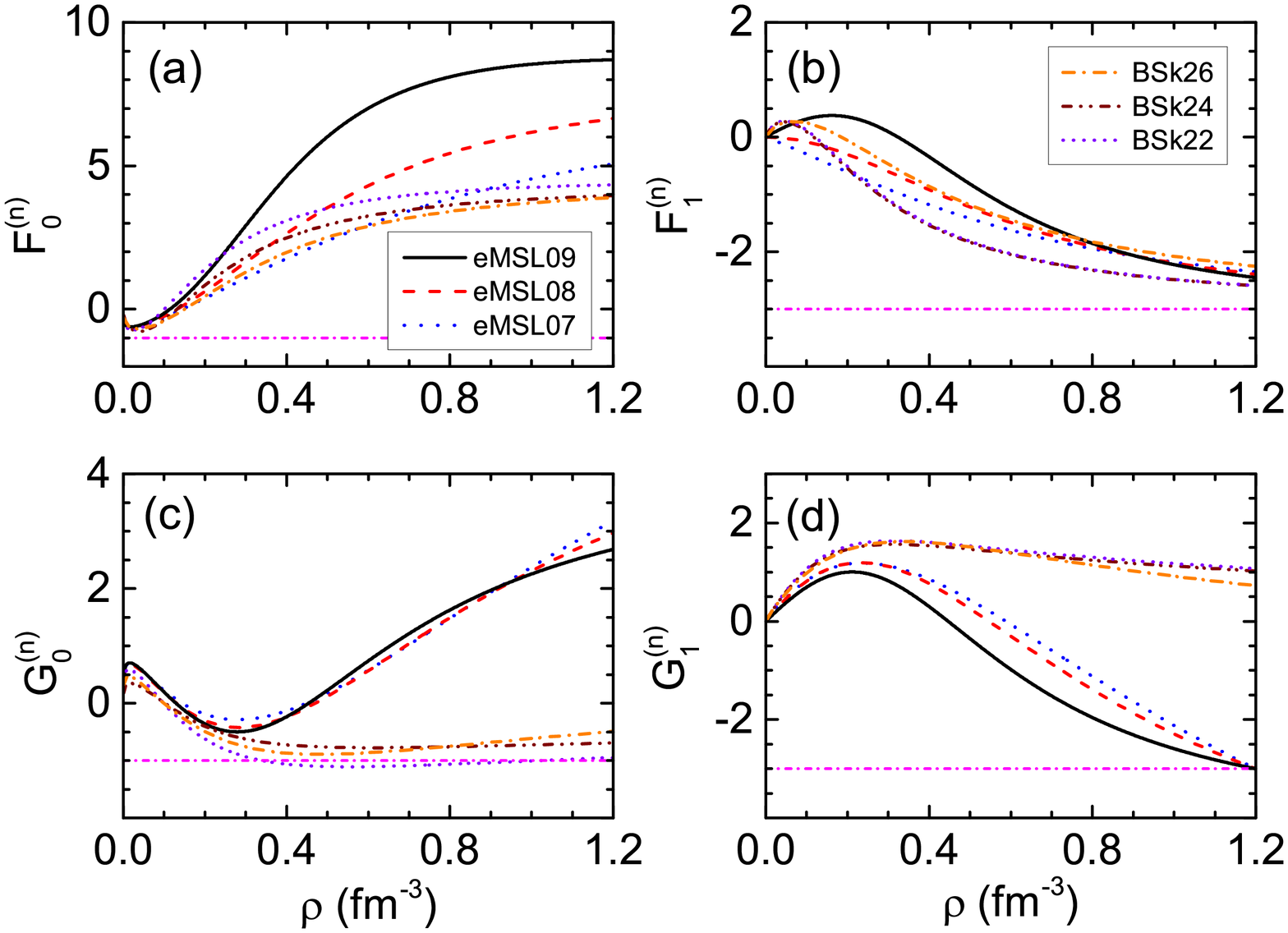}
\caption{(Color online) Similar to Fig.~\ref{Fig:LPsnm} but
for pure neutron matter.}
\label{Fig:LPpnm}
\end{figure}

We calculate the density dependence of the Landau parameters of symmetric nuclear matter and
pure neutron matter according to the explicit expressions in Refs.~\cite{Cha09,Gor10}
with the eMSL and BSk family, and
the results are exhibited in Fig.~\ref{Fig:LPsnm} and Fig.~\ref{Fig:LPpnm},
respectively. One can see that for the three new extended Skyrme interactions,
all the Landau parameters, except $F_0$ for symmetric nuclear matter, satisfy the stability
conditions at densities up to $1.2$ fm$^{-3}$ and thus guarantee the
stability of symmetric nuclear matter and pure neutron matter from subsaturation densities to very high densities.
The instability of symmetric nuclear matter at densities below about $0.1~\mathrm{fm}^{-3}$
determined by the value of $F_0$ corresponds to the well-known spinodal
instability, which is physical and is believed to be related to the
liquid-gas phase transition in nuclear matter and the multifragmentation
phenomenon observed in heavy-ion collisions at intermediate energies~\cite{Sur89,Zha95}.
In the spinodal instability region of symmetric nuclear matter, the squared sound velocity or the
incompressibility of symmetric nuclear matter is negative (see, e.g., Eq.~(\ref{Eq:Sound}) and
the relation $mv_s^2\approx K(\rho)/9$).
At densities beyond $1.2~\mathrm{fm}^{-3}$, both symmetric nuclear matter and pure neutron matter become unstable
in the spin-isospin channels, and the density $\rho =1.2~\mathrm{fm}^{-3}$ is thus
the critical density for the eMSL interactions.
For BSk24 and BSk26 interactions,
these instabilities are eliminated at even higher densities.
For the interaction BSk22,
the instabilities are also eliminated at very higher densities
except that the ferromagnetic instability in pure neutron matter
will appear in the density region from $0.34$ fm$^{-3}$ to $1.02$ fm$^{-3}$
where one has $G^{(n)}_0 < -1$.
The large value of
$1.2~\mathrm{fm}^{-3}$ of the critical density ensures that the three new
extended Skyrme interactions obtained in the present work can be used
safely to calculate the structure of neutron stars. As we will see later,
the center densities of the neutron stars with largest mass configuration
obtained with eMSL07, eMSL08 and eMSL09 are all less than
$\rho =1.2~\mathrm{fm}^{-3}$.

To close this subsection, we list in Tab.~\ref{Tab:LP} the values of all the
twelve Landau parameters of symmetric nuclear matter and pure neutron matter
at saturation density in the SHF model with the eMSL and BSk family.

\begin{table}
\caption{Landau parameters of symmetric nuclear matter and pure neutron matter at
saturation density in the SHF model with eMSL07, eMSL08, eMSL09, BSk22, BSk24 and BSk26. }
\label{Tab:LP}
\begin{tabular}{lcccccc}
\hline
\hline
               &eMSL07 & eMSL08 & eMSL09  & BSk22 & BSk24  & BSk26   \\
 \hline											
$F_0$          &-0.27  &	-0.17 &	-0.06 & -0.10 & -0.10  & -0.12  \\
$F_0^{\prime}$ &0.88   &	1.15  &	1.42  &  1.10 &  0.97  &  0.96  \\
$F_1$          &-0.90  &	-0.60 &	-0.30 & -0.60 & -0.60  & -0.60  \\
$F_1^{\prime}$ &0.50   &	0.43  & 0.60  &  0.40 &  0.37  &  0.67  \\
$G_0$          & 0.21  &	0.16  &	0.05  & -0.24 & -0.061 & -0.21  \\
$G_0^{\prime}$ &0.24   &	0.19  &	0.17  &  0.39 &  0.40  &  0.41  \\
$G_1$          & 0.22  &	0.17  &	-0.08 &	 0.64 &	 0.63  &  0.48  \\
$G_1^{\prime}$ &0.67   &	0.76  &	0.83  &  0.56 &  0.55  &  0.54  \\
$F_0^{(n)}$    &0.09   &	0.23  &	0.54  &  0.83 &  0.32  & -0.074  \\
$F_1^{(n)}$    &-0.48  &    -0.20 &	0.37  & -0.23 & -0.27  & -0.094  \\
$G_0^{(n)}$    &-0.06  &	-0.14 &	-0.17 & -0.41 & -0.26  & -0.32  \\
$G_1^{(n)}$    &1.07   &	1.08  &	0.93  &  1.40 &  1.34  &  1.31 \\
\hline
\hline
\end{tabular}
\end{table}

\subsection{Neutron star properties}

In the following, we discuss the mass-radius relation of static neutron
stars, which is obtained by solving the famous Tolman-Oppenheimer-Volkov
(TOV) equation~\cite{Tol39,Opp39}, i.e.,
\begin{eqnarray}
\frac{dP(r)}{dr}&=&-\frac{G}{c^2r^2}\left[\epsilon(r)+\frac{P(r)}{c^2}\right] \notag\\
&\times & \left[M(r)+4\pi r^3 \frac{P(r)}{c^2} \right]
\left[1-\frac{2GM(r)}{c^2r}\right]^{-1}, \\
\frac{dM(r)}{dr}&=&4\pi\epsilon(r) r^2/c^2,
\end{eqnarray}
where $r$ is the radial coordinate, $M(r)$ is the gravitational mass
inside the sphere of radius $r$, $\epsilon(r)$ and $P(r)$ are,
respectively, the corresponding energy density and pressure of the
neutron star matter at $r$, and $G$ is Newton's gravitational constant,

To solve the TOV equation, an EOS of neutron star matter $P(\epsilon)$ is
necessary. In the present work, we assume the core of neutron stars
consists of neutron, proton, electrons and possible muons without phase
transition and other degrees of freedom at high densities. Then the EOS
of neutron star matter is constructed in the following way: for
the core, we calculate the EOS of $npe\mu$ matter in SHF model using the
extended Skyrme interactions; for the outer crust, we use the EOS of BPS (FMT)
in the region of $6.93\times 10^{-13}~\mathrm{fm}^{-13} < \rho < \rho_{\mathrm{out}}$
($4.73\times 10^{-15}$~fm $^{-3} < \rho < 6.93\times10^{-13}~\mathrm{fm}^{-3}$ )~\cite{Bay71};
for the inner crust in the density region between $\rho_{\mathrm{out}}$
and $\rho_{\mathrm{t}}$ we construct its EOS by interpolation
with the form~\cite{Car03,Xu09}
\begin{equation}
P = a + b \epsilon ^{4/3}.
\end{equation}
In this work, the critical density between the inner and the outer
crust is taken to be $\rho_{\mathrm{out}}=2.46\times10^{-4}~\mathrm{fm}^{-3}$~\cite{Car03,Xu09},
and $\rho_t$ is the core-crust transition density which is evaluated
self-consistently using the extended Skyrme interactions within the
thermodynamic method (see, e.g., Ref.~\cite{Xu09}).

\begin{table}
\caption{Maximum mass of the neutron star ($M_{\mathrm{max}}$), the center
density of the maximum mass neutron star configuration ($\rho^{\text {cen}}_{\text {max}}$),
the radius of $1.4M_{\odot}$ neutron star ($R_{1.4}$), and core-crust
transition density of the neutron star ($\rho_t$) for eMSL07, eMSL08, eMSL09, BSk22, BSk24 and BSk26.}
\label{Tab:NS}
\begin{tabular}{lcccccc}
\hline
\hline
 & eMSL07 & eMSL08 & eMSL09 & BSk22& BSk24 & Bsk26\\
\hline
$M_{\mathrm{max}}/M_{\odot}$ & 2.17 & 2.19 & 2.21 &2.27& 2.28 & 2.17\\
$\rho^{\text {cen}}_{\text {max}}$ (fm$^{-3}$)& 1.11 & 1.06 & 1.03&0.98 &0.98 &1.13 \\
$R_{1.4}$ (km) & 12.3 & 12.5 & 12.7 & 13.2&12.5&11.8 \\
$\rho_t$ (fm$^{-3}$)&0.076 &0.077 &0.078 & 0.071&0.080&0.085\\
\hline
\hline
\end{tabular}
\end{table}

\begin{figure}[!hbp]
\includegraphics[width=0.9\linewidth]{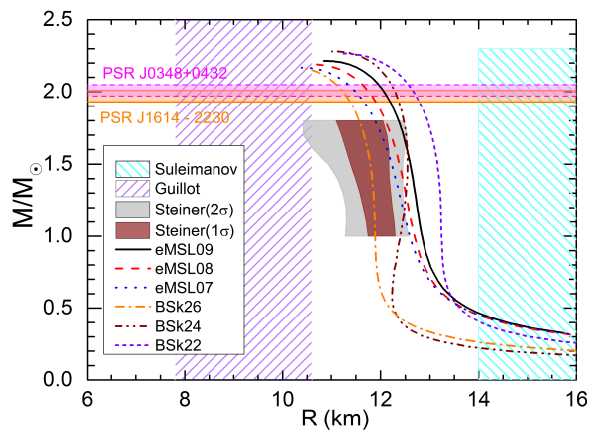}
\caption{(Color online) Mass-radius relation of neutron stars obtained with
eMSL07, eMSL08, eMSL09, BSk22, BSk24 and BSk26. Some recent observational
constraints~\cite{Dem10,Ant13,Ste10,Sul11,Gui13}
are also included for comparison (see text for the details).}
\label{Fig:MR}
\end{figure}

We present in Tab.~\ref{Tab:NS} the maximum mass $M_{\mathrm{max}}$,
the center density $\rho^{\text {cen}}_{\text {max}}$ of the maximum mass
neutron star configuration, the radius of $1.4M_{\odot}$ neutron star
$R_{1.4}$ and the core-crust transition density $\rho_{t}$
for eMSL07, eMSL08, eMSL09, BSk22, BSk24 and BSk26~\cite{Gor13}.
It can be seen that all the six extended Skyrme interactions can
successfully support $2M_{\odot}$ neutron stars.
From the results of the eMSL family, one can find that
the interaction with a stiffer symmetry energy at high-density region
predicts a larger maximum mass and a larger stellar radius for a
canonical neutron star with mass of $1.4M_{\odot}$.
In addition, as mentioned earlier, the
values of $\rho^{\text {cen}}_{\text {max}}$ for the three new extended
Skyrme interactions are all less than the critical density of
$\rho_{\text {cr}} =1.2~\mathrm{fm}^{-3}$. For the core-crust transition
density $\rho_t$, which is strongly correlated with $L(\rho_c)$ or
$L(\rho_0)$~\cite{Che10,Zha14}, the eMSL interactions
also produce very similar values that are consistent with the empirical
value~\cite{Lat04} as well as the result with BSk interactions.

Shown in Fig.~\ref{Fig:MR} is the mass-radius relation of static
neutron stars obtained with eMSL09, eMSL08, eMSL07, BSk22, BSk24 and BSk26~\cite{Gor13}.
It is interesting to see that the BSk24 interaction
predicts smaller radii for low-mass neutron stars
while larger radii for high-mass neutron stars.
This could be due to
the fact that the BSk24 interaction predicts a softer symmetry energy around
saturation density while stiffer SNM EOS  and symmetry energy at supra-saturation
densities.
The horizontal bands in Fig.~\ref{Fig:MR} indicate the measured masses
$M=1.97\pm0.04M_{\odot}$~\cite{Dem10} and $M=2.01\pm0.04M_{\odot}$~\cite{Ant13}
of the two heaviest neutron stars PSR J1614-2230~\cite{Dem10} and
PSR J0348+0432~\cite{Ant13}, respectively.
For comparison, we also show in Fig.~\ref{Fig:MR} the constraints
on the radii of neutron stars by Steiner \textit{et al.} from analyses
of three X-ray bursters and three transient low-mass X-ray binaries~\cite{Ste10}.
It should be noted that, by analyzing the same data sets using more sophisticated
atmosphere models, Suleimanov \textit{et al.} concluded that the lower
limit on the stellar radius is $14$ km for masses below
$2.3M_{\odot}$~\cite{Sul11}, which is included in Fig.~\ref{Fig:MR} 
for comparison.
Indeed, large systematic uncertainties in the analysis of X-ray bursters
have hindered the reliability of these results. Another way to extract the
radii of neutron stars is from the observation of quiescent low mass
X-ray binaries (qLMXB) in globular clusters. By assuming the
neutron star radius is independent on mass, namely $R(M)=R_0$,
Guillet \textit{et al.}~\cite{Gui13} determined a rather small stellar
radius of $R_0 = 9.1^{+1.3}_{-1.5} \mathrm{km}$
within $90\%$ confidence level from fitting the spectra of five
qLMXBs, which is also displayed in Fig.~\ref{Fig:MR} for comparison.
Note that the assumption $R(M)=R_0$ is rather strong and
the additional conditions of causality together with $2M_{\odot}$ neutron stars may
change the result~\cite{Lat14}.
Therefore, the accurate determination of the radius of neutron stars,
which can put a stringent constraint on the density dependence of the
symmetry energy beyond saturation density~\cite{Lat01}, is still a
big challenge.
The radii of the $1.4M_{\odot}$ neutron star ($\sim12.5$ km)
predicted by the eMSL family are in very good agreement with recent studies based on nuclear experiments and theoretical calculations
of pure neutron matter (see, e.g., Ref.~\cite{Lat14} and references therein).

\section{Conclusions}
\label{Sec:Summary}

Within the framework of the extended Skyrme interaction which
includes additional momentum and density dependent terms to
simulate the momentum dependence of three-body nuclear force,
we have constructed three new extended Skyrme interactions, namely,
eMSL07, eMSL08 and eMSL09, by fitting experimental data of a number of finite nuclei
together with a few additional constraints on nuclear matter with the simulated annealing
method.
We have shown that the eMSL family of extended Skyrme interactions
can reasonably describe the ground-state properties and the isoscalar
giant monopole resonance energies of various spherical nuclei used in the fit
as well as the ground-state properties of many other spherical nuclei,
and conform to various most recent constraints
on nuclear matter EOS from theory, experiment and observation.
The high-density EOSs of
pure neutron matter for the three new extended
Skyrme interactions are also consistent with the predictions of the
microscopic calculations using realistic nuclear forces.
Moreover, the three new extended Skyrme interactions successfully eliminate
the unphysical instabilities in nuclear matter at densities up to
about $7.5\rho_0$, and so the eMSL family constructed in the
present work are suitable for the study of neutron stars.
We have used
the three new extended Skyrme interactions to study
the mass-radius relation of neutron stars and our results indicate
that the eMSL family can support $2M_{\odot}$ neutron stars and
predict very reasonable stellar radii.

We have also made comparison for the predictions of finite nuclei,
nuclear matter and neutron stars with the three new eMSL interactions
versus those with the accurately calibrated interactions BSk22, BSk24 and BSk26.
Although obtained through very different fitting strategies,
the new eMSL interactions and BSk interactions give quite similar predictions
for nuclear matter and neutron stars with the main difference appeared
for some higher-order characteristic parameters of asymmetric nuclear
matter, e.g., $J_0$, $L$, $K_{\text{sym}}$, and $K_{\text{sat,2}}$.
For finite nuclei, while the present eMSL family in Hartree-Fock
calculations with the BCS pairing framework under the constant gap
approximation can reasonably describe
the binding energies and charge rms radii of the ground-state spherical
even-even nuclei as well as the isoscalar giant monopole resonance energies
of several representative finite nuclei,
we have noted that the more sophisticated Bogoliubov
treatment for pairing effects with a microscopic pairing force as well as
some additional corrections, such as the Wigner terms and the spurious
collective energy, can significantly improve the description of
the binding energy over the whole nuclear chart as demonstrated in the
Hartree-Fock-Bogoliubov calculations with BSk22, BSk24 and BSk26.
We would like to point out that
a very accurate nuclear mass model over the whole nuclear chart is
beyond the scope of the present work and perhaps this could be pursued
in future.

In conclusion, by including  the additional
momentum and density dependent two-body terms in the Skyrme interaction,
which effectively simulate the momentum dependence of
three-body nuclear force,  we have constructed three new extended
Skyrme interactions, namely,
eMSL07, eMSL08 and eMSL09, which can simultaneously well describe
nuclear matter, finite nucle and neutron stars.
The three new eMSL interactions
predict very similar properties of
finite nuclei and nuclear matter below and around saturation density
but different high-density behaviors of the symmetry energy, and thus
they are potentially useful for the study of the currently largely
uncertain high-density symmetry energy.

\begin{acknowledgments}
This work was supported in part by the Major State Basic
Research Development Program (973 Program) in China under Contract Nos.
2013CB834405 and 2015CB856904, the NNSF of China under Grant Nos. 11625521, 11275125
and 11135011, the ``Shu Guang" project supported by Shanghai Municipal
Education Commission and Shanghai Education Development
Foundation, the Program for Professor of Special Appointment (Eastern Scholar)
at Shanghai Institutions of Higher Learning, and the Science and Technology
Commission of Shanghai Municipality (11DZ2260700).
\end{acknowledgments}

\appendix
\section{Macroscopic quantities in the extended SHF model}
\label{App:EOSESHF}

In the SHF model with the extended Skyrme interaction given in
Eqs.~(\ref{Eq:Sky}) and (\ref{Eq:ExSky}), the EOS of asymmetric
nuclear matter can be expressed as~\cite{Cha09}
\begin{eqnarray}
\label{Eq:EANM}
E(\rho,\delta)&=&\frac{3\hbar^2}{10m}k_F^2F_{5/3} \notag \\
 &+&\frac{1}{8}t_0\rho[2(x_0+2)-(2x_0+1)F_2]\notag\\
 &+&\frac{1}{48}t_3{\rho}^{\alpha+1}[2(x_3+2)-(2x_3+1)F_2]\notag \\
 &+&\frac{3}{40}\rho k_F^2 \Big\{ [t_1(x_1+2)+t_2(x_2+2)]F_{5/3} \notag \\
 &+&\frac{1}{2}[t_2(2x_2+1)-t_1(2x_1+1)]F_{8/3}\Big\}\notag\\
 &+&\frac{3}{40}\rho k_F^2 \Big\{ [t_4(x_4+2)\rho^{\beta}+t_5(x_5+2)\rho^{\gamma}]F_{5/3} \notag \\
 &+&\frac{1}{2}[t_5(2x_5+1)\rho^{\gamma}-t_4(2x_4+1)\rho^{\beta}]F_{8/3}\Big\},
\end{eqnarray}
where $m$ is the nucleon mass, $k_F$ is the Fermi momentum of symmetric
nuclear matter, i.e.,
\begin{equation}
k_F =\left( \frac{3\pi ^2}{2}\rho \right)^{1/3},
\end{equation}
and $F_x(\delta)$ is expressed as
\begin{equation}
  F_x(\delta)=\frac{1}{2}[(1+\delta)^x+(1-\delta)^x].
\end{equation}
By setting $\delta=0$ in Eq.~(\ref{Eq:EANM}), one can obtain the EOS of
symmetric nuclear matter as~\cite{Cha09}
\begin{eqnarray}
\label{Eq:E0}
E_0(\rho)&=
&\frac{3\hbar^2}{10m}k_{F}^2+\frac{3}{8}t_0\rho+\frac{1}{16}t_3\rho^{\alpha+1} \notag \\
&+&\frac{3}{80}[3t_1+t_2(4x_2+5)]{\rho}k_{F}^2 \notag\\
&+&\frac{3}{80}[3t_4\rho^{\beta}+t_5(4x_5+5)\rho^{\gamma}]{\rho}k_{F}^2.
\end{eqnarray}
The incompressibility $K_0$ and skewness parameter
$J_0$ of symmetric nuclear matter can then be easily
derived with the following definitions
\begin{eqnarray}
K_0 &\equiv& 9\rho_0^2\left.\frac{d^2 E_0(\rho)}{\partial \rho^2}\right\vert_{\rho=\rho_0}, \\
J_0& \equiv&27\rho_0^3\left.\frac{d^3 E_0(\rho)}{\partial \rho^3}\right\vert_{\rho=\rho_0}.
\end{eqnarray}
The symmetry energy is given by~\cite{Cha09}
\begin{eqnarray}
\label{Eq:SE}
E_{\mathrm{sym}}(\rho)& \equiv&\frac{1}{2}\left.\frac{\partial ^2 E(\rho,\delta)}{\partial \delta ^2}\right\vert_{\delta=0} \notag \\
&=&\frac{\hbar^2}{6m}k_{F}^2-\frac{1}{8}t_0(2x_0+1)\rho \notag \\
&-&\frac{1}{48}t_3(2x_3+1)\rho^{\alpha+1}\notag\\
&-&\frac{1}{24}[3t_1x_1 -t_2(4+5x_2)]\rho k_{F}^2\notag\\
&-&\frac{1}{24}[3t_4x_4\rho^{\beta} -t_5(4+5x_5)\rho^{\gamma}]\rho k_{F}^2,
\end{eqnarray}
and the density slope of the symmetry energy at
a reference density $\rho_r$ can be obtained by the following definition
\begin{equation}
L(\rho_r)\equiv 3\rho_r\left.\frac{d E_{\mathrm{sym}}(\rho)}{d\rho}\right\vert_{\rho=\rho_r}.
\end{equation}

In the extended SHF model, the isoscalar effective mass $m_{s,0}^*$,
namely, the effective mass of nucleon in symmetric nuclear matter at
saturation density is evaluated by the following relation~\cite{Cha09}
\begin{eqnarray}
\frac{\hbar^2}{2m_{s,0}^*}&=&\frac{\hbar^2}{2m}+\frac{3}{16}t_1\rho_0+\frac{1}{16}t_2(4x_2+5)\rho_0 \notag\\
&+&\frac{3}{16}t_4\rho_0^{\beta+1}+\frac{1}{16}t_5(4x_5+5)\rho_0^{\gamma+1}.
\end{eqnarray}
The value of the isovector effective mass $m_{v,0}^*$, namely,
the  effective mass of neutron (proton) in pure proton (neutron) matter
at saturation density is given by~\cite{Cha09}
\begin{eqnarray}
\frac{\hbar^2}{2m_{v,0}^*}&=&\frac{\hbar^2}{2m}+\frac{1}{8}t_1(x_1+2)\rho_0+\frac{1}{8}t_2(x_2+2)\rho_0\notag\\
&+&\frac{1}{8}t_4(x_4+2)\rho_0^{\beta+1}+\frac{1}{8}t_5(x_5+2)\rho_0^{\gamma+1}.
\end{eqnarray}

For the finite-range term $\mathcal{H}_{\mathrm{fin}}$ in
Eq.~(\ref{Eq:HESkyrme}), the gradient coefficient $G_S$,
the symmetry gradient coefficient $G_V$ and the cross gradient
coefficient $G_{SV}$ can be obtained as
\begin{eqnarray}
G_S& =&\frac{9}{32}t_1-\frac{1}{32}t_2(4x_2+5)\notag\\
    &+&\frac{3}{32}t_4(2\beta+3)\rho^{\beta}-\frac{1}{32}t_5(4x_5+5)\rho^\gamma,\\
G_V &=&\frac{3}{32}t_1(2x_1+1)+\frac{1}{32}t_2(2x_2+1) \notag\\
    &+&\frac{3}{32}t_4(2x_4+1)\rho^{\beta}+\frac{1}{32}t_5(2x_5+1)\rho^{\gamma},\\
G_{SV}&=&\frac{\beta}{16}t_4(1+2x_4)\rho^{\beta}.
\end{eqnarray}

\section{Relationship between the Skyrme parameters and the macroscopic
quantities in the extended SHF model}
\label{App:MSL}

We reexpress the EOS of symmetric nuclear matter (Eq.~(\ref{Eq:E0})) as
\begin{eqnarray}
E_0(\rho)&=&E_{\mathrm{kin}}^0(\frac{\rho}{\rho_0})^{2/3}+s_0\frac{\rho}{\rho_0}+(s_1+s_2)(\frac{\rho}{\rho_0})^{5/3} \notag \\
&+&s_3(\frac{\rho}{\rho_0})^{\alpha+1}+s_4(\frac{\rho}{\rho_0})^{\beta+5/3}+s_5(\frac{\rho}{\rho_0})^{\gamma+5/3},
\end{eqnarray}
where we have
\begin{equation}
E_{\mathrm{kin}}^0=\frac{3\hbar^2}{10m}k_{F,0}^2,
\end{equation}
with $k_{F,0}=(3\pi^2\rho_0/2)^{1/3}$ and the coefficients
$s_0\sim s_5$ are defined as
\begin{align}
&s_0=\frac{3}{8}t_0\rho_0, & &s_3=\frac{1}{16}t_3\rho_0^{\alpha+1}, \notag\\
&s_1=\frac{9}{80}t_1\rho_0 k_{F,0}^2,& &s_4=\frac{9}{80}t_4\rho_0^{\beta+1} k_{F,0}^2 \notag \\
&s_2 = \frac{3}{80}t_2(5+4x_2)\rho_0 k_{F,0}^2,& &
s_5=\frac{3}{80}t_5(5+4x_5)\rho_0^{\gamma+1} k_{F,0}^2 \notag
\end{align}
Similarly, the symmetry energy (Eq.~(\ref{Eq:SE})) can be rewritten as
\begin{eqnarray}
E_{\mathrm{sym}}(\rho)&=&E_{\mathrm{kin}}^{\mathrm{sym}}(\frac{\rho}{\rho_0})^{2/3}+\omega_0\frac{\rho}{\rho_0}+
(\omega_1+\omega_2)(\frac{\rho}{\rho_0})^{5/3}\notag\\
&+&\omega_3(\frac{\rho}{\rho_0})^{\alpha+1}+\omega_4(\frac{\rho}{\rho_0})^{\beta+5/3}+\omega_5(\frac{\rho}{\rho_0})^{\gamma+5/3},
\end{eqnarray}
where we have
\begin{equation}
E_{\mathrm{kin}}^{\mathrm{sym}}=\frac{\hbar^2}{6m}k_{F,0}^2
\end{equation}
and the coefficients $\omega_0 \sim \omega_5$ are defined as
\begin{align}
&\omega_0=-\frac{1}{8}t_0(2x_0+1)\rho_0, &
& \omega_3=\frac{1}{48}t_3(2x_3+1)\rho_0^{\alpha+1},\notag \\
&\omega_1=-\frac{1}{8}t_1x_1\rho_0 k_{F,0}^2, &
&\omega_4=-\frac{1}{8}t_4x_4\rho_0^{\beta+1} k_{F,0}^2,\notag\\
&\omega_2=\frac{1}{24}t_2(4+5x_2)\rho_0 k_{F,0}^2, &
&\omega_5=\frac{1}{24}t_5(4+5x_5)\rho_0^{\gamma+1} k_{F,0}^2. \notag
\end{align}

In the extended SHF model with fixed $\beta$ and $\gamma$,
the thirteen coefficients $s_0\sim s_5$, $\omega_0 \sim \omega_5$ and
$\alpha$ can be explicitly expressed in terms of thirteen macroscopic
quantities $\rho_0$, $E_0(\rho_0)$, $K_0$, $J_0$, $E_{\mathrm{sym}}(\rho_r)$,
$L(\rho_r)$, $K_{\mathrm{sym}}(\rho_r)$, $G_S$, $G_V$, $G_{SV}$,
$G_0^{\prime}$, $m_{s,0}^{\ast}$ and $m_{v,0}^{\ast}$.
Before showing the explicit expressions, we define
\begin{align}
& \xi_1 = E_{\mathrm{kin}}^{0}\frac{m-m_{s,0}^{\ast}}{m_{s,0}^{\ast}}, &&
\xi_2 = \frac{20}{9}\xi_1-\frac{\hbar^2}{2m}\frac{m-m_{v,0}^{\ast}}{m_{v,0}^{\ast}}k_{F,0}^2, \notag \\
&A_0^{\prime}=\frac{27\hbar ^2 \pi ^2G_0^{\prime}\rho_0}{4m_{s,0}^{\ast}k_{F,0}},& &
A_S=\frac{6}{5}G_{S}\rho_0k_{F,0}^2 \notag\\
&A_V=18G_{V}\rho_0k_{F,0}^2,& &
A_{SV}=\frac{1}{\beta}{G_{SV}\rho_0k_{F,0}^2}. \notag
\end{align}
The $s_0\sim s_5$, $\omega_0 \sim \omega_5$ and $\alpha$ can then be
expressed as
\begin{eqnarray*}
s_4&=&\frac{1}{2\beta}\left(A_S+4Y_1-3\xi_1 \right), \\
s_1&=&\xi_1-Y_1-s_4, \\
\alpha &=& \frac{-Y_b+\sqrt{Y_b^2-4Y_a\cdot Y_c}}{2Y_a},\\
s_3& = &\frac{27Y_a}{\alpha(3\alpha-3\gamma-4)},\\
s_5& = &[K_0+2E_{\mathrm{kin}}^0-10\xi_1-(3\alpha+3)
(E_{\mathrm{kin}}-3E_0-2\xi_1) \notag \\
& &-3\beta(3\beta-3\alpha+4)s_4
]/\left[ 3\gamma(3\gamma-3\alpha+4)\right], \\
s_0 &=& E_0-E_{\mathrm{kin}}^0-\xi_1-s_3, \\
s_2 &=& Y_1-s_5,
\end{eqnarray*}

\begin{eqnarray*}
\omega_4 &=& \frac{5}{9}s_4-A_{SV}, \\
\omega_1 &=& -\omega_4-\frac{A_0^{\prime}}{54}-\frac{A_V}{27}
-\frac{E_0}{6}+\frac{E_{\mathrm{kin}}^0}{6}
+\frac{4}{9}\xi_1, \\
\omega_3 &=&\Big\lbrace Y_2\left[(3\gamma+2)(3\gamma+5)\eta ^{\gamma+5/3}-10\eta^{5/3}\right] \\
&&-Y_3\left[(3\gamma+2)\eta ^{\gamma+5/3}-2\eta ^{5/3} \right] \Big\rbrace \notag \\
& &/ \Big\lbrace 3\alpha\eta^{\alpha+1}\big[(3\gamma+2)(3\gamma-3\alpha+2)\eta^{\gamma+5/3} \\
& & +(6\alpha-4)\eta^{5/3}
\big] \Big\rbrace, \\
\omega_5&=&[Y_3-Y_2(3\alpha+3)] \\
&&/
\left[(3\gamma+2)(3\gamma-3\alpha+2)\eta^{\gamma+5/3}
+2(3\alpha-2)\eta ^{5/3}\right], \\
\omega_2&=&-\frac{A_0^{\prime}}{18}-\frac{E_0}{2}+\frac{E_{\mathrm{kin}}^0}{2}-\frac{\xi_1}{3}+\frac{20}{9}Y_1-\omega_5, \\
\omega_0&=&{\eta}^{-1}{E_{\mathrm{sym}}(\rho_r)}
-\eta^{-1/3}E_{\mathrm{kin}}^{\mathrm{sym}}
-\eta^{\alpha}\omega_3 \\
&& -\eta^{2/3}(\omega_1+\omega_2)-\eta^{\beta+2/3}\omega_4
-\eta^{\gamma+2/3}\omega_5,
\end{eqnarray*}
with
\begin{eqnarray*}
\eta & = & \frac{\rho_r}{\rho_0}, \\
Y_1&=&\frac{1}{30}A_0^{\prime}+\frac{1}{60}A_V
+\frac{3}{10}E_0-\frac{3}{10}E_{\mathrm{kin}}^0
-\frac{1}{20}\xi_1+\frac{9}{20}\xi_2, \\
Y_2&=&L(\rho_r)-3E_{\mathrm{sym}}(\rho_r)
+\eta ^{2/3}E_{\mathrm{kin}}^{\mathrm{sym}}
-2\eta^{5/3}\xi_2 \\
& &+\omega_4[2\eta^{5/3}-(3\beta+2)\eta^{\beta+5/3}], \\
Y_3&=&K_{\mathrm{sym}}(\rho_r)
+2\eta^{2/3}E_{\mathrm{kin}}^{\mathrm{sym}}
-10\eta^{5/3}\xi_2+ \\
& &\omega_4\left[10\eta^{5/3}
-(3\beta+2)(3\beta+5)\eta^{\beta+5/3}\right], \\
Y_a&=&9\left[ K_0+2E_{\mathrm{kin}}^0
-10\xi_1-(3\gamma+7)(E_{\mathrm{kin}}^0
-3E_0-2\xi_1) \right.\\
& & \left.-9\beta(\beta-\gamma)s_4 \right], \\
Y_b&=&9(3\gamma ^2+6\gamma+1)
(E_{\mathrm{kin}}^0-3E_0-2\xi_1-3\beta s_4)\notag\\
& &-3\left[J_0-8E_{\mathrm{kin}}^0+10\xi_1
-9\beta(3\beta ^2+6\beta+1)s_4\right], \\
Y_c&=&-Y_a-3(3\gamma^2+6\gamma+1) \\
& & \times\left[K_0-E_{\mathrm{kin}}^0+9E_0
-4\xi_1-3\beta (3\beta+4)s_4 \right] \notag \\
&& +(3\gamma+4)\left[J_0-8E_{\mathrm{kin}}^0+10\xi_1
-9\beta(3\beta^2+6\beta+1)s_4\right].
\end{eqnarray*}
Once given the thirteen macroscopic quantities, one
can obtain $s_0\sim s_5$, $\omega_0 \sim \omega_5$ and
$\alpha$ by invoking the above expressions, and
then the Skyrme parameters can be easily obtained.

\end{document}